\documentclass[12pt]{article}
\usepackage{amssymb}
\usepackage{amsmath}
\usepackage{bm}
\usepackage{a4}
\usepackage{graphicx}
\usepackage{comment}

\DeclareMathSymbol{\mrq}{\mathord}{operators}{`'}

\evensidemargin \oddsidemargin
\newcommand{\RR}{\mathbb{R}}

\def\1{{\bf 1}}
\def\0{{\bf 0}}

\def \D {{\cal D}}

\def \I {{\cal I}}

\def \U {{\cal U}}

\newcommand{\Be}{{\bm \epsilon}}
\newcommand{\Bep}{{\bm \epsilon}^{\scriptscriptstyle \perp}}

\newcommand{\bi}{\mathbf{i}}
\newcommand{\bj}{\mathbf{j}}
\newcommand{\bk}{\mathbf{k}}

\def\nn{\nonumber \\}

\newcommand{\bx}{{\bm x}}

\newcommand{\bX}{{\bm X}}
\newcommand{\bXp}{{\bm X}^{\scriptscriptstyle \perp}}

\newcommand{\bv}{{\bm v}}

\newcommand{\Bap}{{\bm \alpha}^{\scriptscriptstyle \perp}}

\newcommand{\bE}{{\bm E}}
\newcommand{\bEp}{{\bm E}^{\scriptscriptstyle \perp}}
\newcommand{\bB}{{\bm B}}

\newcommand{\hD}{\Delta}
\newcommand{\hs}{ s}
\newcommand{\hH}{ H}
\newcommand{\hJ}{ J}
\newcommand{\hsg}{ \sigma}
\newcommand{\hbx}{\bx}
\newcommand{\hxe}{ x_e}
\newcommand{\hye}{ y_e}
\newcommand{\hze}{ z_e}
\newcommand{\hp}{ p}
\newcommand{\hbp}{{\bm p}}
\newcommand{\hbu}{{\bm u}}
\newcommand{\hDO}{\Delta\!^{{\scriptscriptstyle (0)}}}
\newcommand{\hDU}{\Delta\!^{{\scriptscriptstyle (1)}}}
\newcommand{\hDD}{\Delta\!^{{\scriptscriptstyle (2)}}}
\newcommand{\hDOu}{\Delta_u}
\newcommand{\hDd}{\Delta_d}

\newcommand{\vM}{v_{{\scriptscriptstyle M}}}
\newcommand{\tvM}{\tilde v_{{\scriptscriptstyle M}}}
\newcommand{\xiH}{\xi_{{\scriptscriptstyle H}}}

\newcommand{\sU}{s^{{\scriptscriptstyle (1)}}}
\newcommand{\sD}{s^{{\scriptscriptstyle (2)}}}
\newcommand{\hsU}{\hs ^{{\scriptscriptstyle (1)}}}
\newcommand{\hsD}{\hs ^{{\scriptscriptstyle (2)}}}
\newcommand{\Hu}{H_{{\scriptscriptstyle u}}}
\newcommand{\Hd}{H_{{\scriptscriptstyle d}}}

\newcommand{\be}{\begin{equation}}
\newcommand{\ee}{\end{equation}}
\newcommand{\bea}{\begin{eqnarray}}
\newcommand{\eea}{\end{eqnarray}}
\newcommand{\ba}{\begin{array}}
\newcommand{\ea}{\end{array}}
\newtheorem{prop}{Proposition}
\newtheorem{lemma}{Lemma}
\newtheorem{theorem}{Theorem}
\newtheorem{corollary}{Corollary}
\newtheorem{Definition}{Definition}
\def\sq{\mbox{\rlap{$\sqcap$}$\sqcup$}}
\newenvironment{proof}[1]{\vspace{5pt}\noindent{\bf Proof #1}\hspace{6pt}}%
{\hfill\sq}
\newcommand{\bp}{\begin{proof}}
\newcommand{\ep}{\end{proof}\par\vspace{10pt}\noindent}
\begin{document}

\title{Hydrodynamic impacts of short laser pulses on plasmas}

\author{Gaetano Fiore$^{1,3}$\footnote{Corresponding author. Email: gaetano.fiore@na.infn.it}, \ Monica De Angelis$^{1}$, Renato Fedele$^{2,3}$, \\
 Gabriele Guerriero$^{1}$, Du\v{s}an Jovanovi\'c$^{4,5}$, %Tahmina Akhter$^{1,3}$, Sergio De Nicola
   \\    %  \and
$^{1}$ Dip. di Matematica e Applicazioni, Universit\`a di Napoli ``Federico II'',\\
$^{2}$  Dip. di Fisica, Universit\`a di Napoli ``Federico II'', \\% Complesso  MSA,  Via Cintia, 80126 Napoli, Italy
Complesso Universitario  M. S. Angelo, Via Cintia, 80126 Napoli, Italy\\         %\and
$^{3}$         INFN, Sez. di Napoli, Complesso  MSA,  Via Cintia, 80126 Napoli, Italy\\ 
$^{4}$  Inst. of Physics, University of Belgrade, %Pregrevica 118, 
11080 Belgrade, Serbia\\
$^{5}$  Texas A \& M University at Qatar, 23874 Doha, Qatar 
}

\date{}
\maketitle

\begin{abstract} 
We determine conditions allowing to simplify the description of the impact of a short and arbitrarily intense laser pulse onto a cold plasma at rest.
If both the initial plasma density and pulse profile have plane simmetry, then suitable matched upper bounds on the maximum and the relative variations of the initial density, as well as 
the intensity and duration of the pulse, ensure a
 strictly hydrodynamic evolution of the electron fluid   (without wave-breaking or vacuum-heating) during its whole interaction with the pulse, while ions can be regarded as immobile. 
   We use a  recently developed fully relativistic plane model whereby the system of the (Lorentz-Maxwell and continuity) PDEs is reduced into a family of highly nonlinear but decoupled systems of non-autonomous Hamilton  
equations with one degree of freedom, with the light-like coordinate $\xi=ct\!-\!z$ instead of time $t$ as an independent variable, and new  apriori estimates (eased by use of a Liapunov function) of the solutions in terms of the input data (initial density and pulse profile). 
If the laser spot radius $R$ is finite but not too small the same conclusions hold for the part of the plasma close to the axis  $\vec{z}$ of cylindrical symmetry. These results
may help in drastically simplifying the study of extreme acceleration mechanisms of electrons.

\end{abstract}

%\noindent
%{\bf Keywords:} \ plasma hydrodynamics; 
%non-autonomous  Hamilton  equations; Liapunov function;
%relativistic electrodynamics; plasma wave; wave-breaking.

\section{Introduction and preliminaries}

Laser-plasma interactions induced by ultra-intense  laser pulses lead to a variety of very interesting phenomena \cite{Kruer19,SprEsaTin90PRA,SprEsaTin90PRL,EsaSchLee09,Mac13}, 
notably plasma compression for inertial fusion \cite{KuzRyz17},  laser wakefield acceleration  (LWFA)  \cite{Tajima-Dawson1979,Sprangle1988,TajNakMou17} and other extremely compact acceleration mechanisms
(e.g. hybrid laser-driven and particle-driven plasma wakefield acceleration \cite{HidEtAl19}) of charged particles, which hopefully will be at the basis of a generation of new, table-top accelerators. This is paramount because accelerators have extremely important applications  in particle physics, materials science, medicine, industry, environmental remediation, etc; therefore huge investments\footnote{We just mention the EU-funded project {\it Eupraxia} \cite{Eupraxia19AIP,Eupraxia19JPCS,Eupraxia20EPJ}.} are being devoted all over the world to the development of such 
%new, table-top 
accelerators. Similar extreme conditions (huge electromagnetic fields and huge accelerations of  charged particles in plasmas) occur also in a number of violent astrophysical processes, see e.g. \cite{TajNakMou17} and references therein.
In general,  these phenomena are ruled by the equations of a kinetic theory coupled to Maxwell equations, which
 can be  solved only numerically via  particle-in-cell (PIC) techniques. 
Unfortunately,  PIC  codes  involve huge and expensive computations
for each choice of the free parameters; even with the presently ever increasing computational
power, exploring the parameter space blindly to single out interesting regions
remains prohibitive. Sometimes,  good predictions can be obtained also by treating the plasma as a multicomponent
(electron and ions) fluid and by numerically solving, via  multifluid codes (such as QFluid \cite{TomEtAl17}) or hybrid kinetic/fluid  codes,  the (simpler) associated hydrodynamic equations; but in general it is not known a priori
in which conditions, or spacetime regions, this is possible. Therefore
all analytical insights that can simplify
the work, at least in special cases or in a limited space-time region, are welcome.

This applies in particular to studying the impact of a very short (and possibly very intense) 
laser pulse perpendicularly onto a cold diluted plasma  at rest 
(or onto matter which is locally  ionized  into a plasma by the front of the pulse itself).
As it is well-known, electrons start oscillating orthogonally to the direction $\vec{z}$ of propagation of the pulse
and drifting in the positive $z$-direction, respectively pushed by  the electric and magnetic parts of the Lorentz force induced by  the pulse; thereafter, electrons  start oscillating also longitudinally (i.e. in $\vec{z}$-direction), pushed by the restoring electric force induced by charge 
separation\footnote{The reader can recognize such initial longitudinal motions e.g. from
fig. \ref{graphs}c and from the electron worldlines reported in fig.s \ref{Worldlines_Nonlin}, \ref{Worldlines_HomLin-Nonlin}.}.
It turns out that the  initial dynamics is simpler if the pulse is
{\it essentially short}. 
We shall say (see Definition \ref{Def1}) that the pulse is {\it essentially short} if it overcomes each electron before the $z$-displacement $\hD$ of the latter  reaches a negative minimum for the first time; that 
an essentially short pulse is {\it strictly short}
if it overcomes each electron before $\hD$ becomes negative for the first time. In other words, we regard a pulse as 
strictly short (resp. essentially short) if it overcomes each  electron before it  finishes the first   $1/2$  (resp. $3/4$) longitudinal oscillation. 
In the nonrelativistic (NR) regime a pulse which is symmetric under inversion about its center is strictly short, essentially short if its duration $l/c$  does not respectively exceed $1/2$, $1$  times the NR plasma oscillation period $t_{{\scriptscriptstyle H}}^{{\scriptscriptstyle nr}}\!\equiv\!\sqrt{\pi m/n_b e^2}$ associated to the maximum $n_b$ of the initial electron density\footnote{If the pulse is  a slowly modulated monochromatic wave (\ref{modulate})
with wavelength $\lambda=2\pi/k$ this implies a fortiori $\frac{4\pi e^2}{mc^2}n_b \lambda^2\ll1$,  so that the plasma is {\it underdense}.}, i.e. if
\be
G_b:=\sqrt{\frac{n_b e^2}{\pi mc^2}}l \: \le \: \left\{\!\!\ba{l} 1/2\\ 1\ea\right. 
% \qquad M_bl^2<....,\quad M_b:=\frac{4\pi e^2}{mc^2}n_b;                        
   \label{Lncond}
\ee
(see Proposition \ref{propshort}); 
here  $-e,m$ are the electron charge and mass, and $c$ is the speed of light.
The {\it relativistic}  plasma oscillation period is not independent of the  oscillation amplitude, but grows with the latter, which in turn grows with the pulse intensity. Correspondingly, 
  eq. (\ref{Lncond'})  can be fulfilled also with a larger $G_b$;
%for which  (\ref{ShortPulse1}),  (\ref{ShortPulse1'}) provide sufficient conditions.
in addition, it is compatible with maximizing the  oscillation amplitude, and thus also the energy transfer from the pulse to the plasma wave, because  for given $n_b$ and pulse energy such a maximization can be achieved \cite{SprEsaTin90PRL,FioFedDeA14} through a suitable
\be
l\sim\tilde\xi_2, \qquad \qquad\mbox{(i.e. when} \:\: G_b\:\sim \: 1/2, \:\:
\mbox{in the nonrelativistic regime)}.            \label{MaxTransfer}
\ee

We believe that such impacts require a deeper understanding because, among other things,  they may 
 generate: i) a plasma wave (PW) \cite{AkhPol56,GorKir1987}, or even a {\it ion bubble}\footnote{Namely, a region containing only ions, because all electrons have been expelled out of it.} \cite{RosBreKat91,MorAnt96,PukMey2002,KosPukKis2004,LuEtAl2006,LuHuaZhoEtAl06,MaslovEtAl16}, producing
the LWFA, i.e.  accelerating  a small bunch of  (socalled {\it witness}) electrons
trailing the pulse  to very high energy,  in the forward direction; ii) the {\it slingshot effect}  \cite{FioFedDeA14,FioDeN16,FioDeN16b},
i.e. the backward acceleration and expulsion of energetic electrons from the vacuum-plasma interface, during or just after the impact.
  The present work is one out of a few papers \cite{FioCat18,Fio-impact%,Fiotdho,FioDisc
} arguing that,  
%As announced in  \cite{FioDeN16},
with the help of the plane, fully relativistic Lagrangian model  of Ref. \cite{Fio14JPA,Fio18JPA} and very little computational power,
 we can obtain  important information about  such an impact, in particular the formation of a PW, its persistence before wave-breaking (WB), the features of the latter. As known, a small WB is not necessarily undesired; it may be used  to produce and inject the mentioned witness electrons in the PW ({\it self-injection}).

\medskip
The plane model is as follows. 
One assumes that the plasma is initially neutral, unmagnetized and at rest with zero densities in the region  \ $z\!<\! 0$.
More precisely, the $t\!=\!0$ initial conditions for  the electron fluid  Eulerian density $n_e$ and velocity $\bv_e$ are of the type
\be 
\bv_e(0 ,\!\bx)\!=\!\0, \qquad n_e(0,\!\bx)\!=\!\widetilde{n_0}(z), 
%\qquad   \qquad j^0(0,\bx) \equiv\sum_h\!q_hn_h(0,\bx) = 0                     
 \label{asyc}
\ee
where the initial electron (as well as proton) density $\widetilde{n_0}(z)$ fulfills
\be 
 \widetilde{n_0}(z)\!=\!0 \:\:\mbox{if }\: z\!\le\! 0, \qquad
0\!<\!\widetilde{n_0}(z)\!\le\! n_b  \quad \mbox{if }\: z\!>\!0
 \label{n_0bounds}
\ee
for some $n_b\!>\!0$  (a few examples are reported in fig. 	\ref{InDensityPlots}). 
%, although we expect that the main results hold also if  $\widetilde{n_0}$ tends to a constant $n_0$ asymptotically as $z\to\infty$. 
%
%As we regard ions as immobile, the proton density will be $ n_p(t,\bx)\!=\!\widetilde{n_0}(z)$ for all $t$.
One assumes that before the impact the laser pulse  is a free plane transverse wave  
%$\Be^{{\scriptscriptstyle\perp}}\!(ct\!-\!z)$ \  
 travelling  in the $z$-direction, i.e.
the electric and magnetic fields $\bE,\bB$ are of the form
\be
\bE (t, \bx)=\bEp (t, \bx)=\Be^{{\scriptscriptstyle\perp}}\!(ct\!-\!z),\qquad  \bB=\bB^{{\scriptscriptstyle\perp}}=
\bk\!\times\!\bEp \qquad \mbox{if } t\le 0             
      \label{pump}
\ee
(the superscript $\perp$ denotes vector components orthogonal to  $\bk\equiv\nabla z$), where
 $\Be^{{\scriptscriptstyle\perp}}\!(\xi)$ has a bounded support
% $S_{\Be^{{\scriptscriptstyle\perp}}}$ 
with $\xi\!=\!0$ as the left extreme (i.e. the pulse reaches the plasma at $t\!=\!0$). 
The input data of a specific problem are the functions $\widetilde{n_0}(z),\Be^{{\scriptscriptstyle\perp}}(\xi)$; it is convenient to define also the related functions
\bea
&\displaystyle \Bap(\xi):= -\!\int^{\xi}_{ -\infty }\!\!\!d\zeta\:\Bep(\zeta), \qquad
&v(\xi):=\left[\frac {e\Bap(\xi)}{mc^2}\right]^2,
%\hbu^{\scriptscriptstyle \perp} (\xi)=-\frac q{mc^2}\Bap(\xi)       
      \label{pump2}\\[8pt]
&\displaystyle \widetilde{N}(Z):=\int^{Z}_0\!\!\! d \zeta\,\widetilde{n_0}(\zeta), \qquad
&\U( \Delta;Z)\!:=\! K\!\!\int^{\Delta}_0\!\!\!\!\!d\zeta\,(\Delta\!-\!\zeta)
\,\widetilde{n_0}(Z\!+\! \zeta)\:,
\eea
where $K:=\frac{4\pi e^2}{mc^2}$.  By definition $v$ is dimensionless and nonnegative, and $\widetilde{N}(Z)$ strictly grows with $Z$.
One describes the plasma as a fully relativistic collisionless fluid of electrons and a static fluid of ions (as usual, in the short time lapse of interest here the motion of the much heavier ions is negligible), 
with  $\bE,\bB$ and the plasma dynamical variables fulfilling the Lorentz-Maxwell and continuity equations. Since at the impact time $t\!=\!0$ the plasma is made of two static fluids, by continuity such a hydrodynamical description (HD) is justified and  one can neglect  the depletion of the pulse  at least for small $t\!>\!0$;  the specific time lapse is %can be
determined   {\it a posteriori},  by self-consistency.  
This allows us to  reduce (see \cite{Fio14JPA,Fio18JPA}, or \cite{Fio14,Fio16b%,Fio17PPLA
,FioCat19,Fio21JPCS} for
 shorter presentations) the system of Lorentz-Maxwell and continuity 
partial differential equations (PDEs)  into ordinary ones, %differential equations % (ODEs), 
more precisely into the following continuous family of   {\it decoupled Hamilton equations for  systems  with one degree of freedom}. Each system determines the complete Lagrangian  (in the sense of non-Eulerian) description of the motion of the electrons having a same
initial longitudinal coordinate $Z>0$ (the {\it $Z$-electrons}, for brevity), and reads
%troppe referenze! distingui contenuti e distribuisci nel testo queste citazioni?
% along the direction of propagation of the pulse. 
%In the model we neglect the motion of ions   and  the depletion of the pulse (this is justified
%for the short time intervals considered here), 
\bea
&\hD '(\xi,Z)%=\hze'(\xi,Z)
=\displaystyle\frac {1\!+\!v(\xi)}{2\hs ^2(\xi,Z)}\!-\!\frac 12, \qquad 
&\hs '(\xi,Z) 
=K\!\left\{\!\widetilde{N}\left[Z\!+\!\hD (\xi,Z)\right] \!-\! \widetilde{N}(Z)\!\right\};
%=\!\!\int^{Z+\hD (\xi,Z)}_Z\!\!\!\!\!\!\!\!\!\!dZ' K\widetilde{n_0}(Z') 
\label{heq1}
\eea
it is  equipped with the initial conditions
\bea
 &\hD (0,Z)=0, \qquad\quad &\hs (0,Z)=1.               \label{incond}
\eea
Here  the unknown basic dynamical variables $\hD (\xi,Z),\hs (\xi,Z)$  are respectivey  the longitudinal displacement and $s$-factor\footnote{Namely, $\hs $ is the light-like component of the 4-velocity  of the $Z$ electrons, or equivalently  is related to their 4-momentum $\hp$
by $\hp^0\!-\!c\hp^z\equiv mc^2 \hs $; it is positive-definite. In the NR regime $|s\!-\!1|\ll1$; in the present fully relativistic regime it needs only satisfy the inequality $s>0$.} of the $Z$-electrons espressed as functions of
 $\xi,Z$, while  $\hze(\xi,Z):=Z\!+\!\hD (\xi,Z)$ is  the present longitudinal coordinate of the $Z$-electrons; 
%the hat $\hat{}$ over the symbol of a dynamical variable means that this is considered as a function of $\xi,Z$ instead of $t,Z$; 
we consider all dynamical variables $f$ (in the Lagrangian description)  
as functions\footnote{In the cited papers \cite{Fio14JPA,Fio18JPA,Fio14,Fio16b%,Fio17PPLA
,FioCat19,Fio21JPCS}  the two dependences are denoted as $\hat f(\xi,Z)$ and $f(t,Z)$ respectively; since here we use only the former, we  denote it simply as $ f(\xi,Z)$ (without the $\hat{}$ ).} of $\xi,Z$ instead of $t,Z$; 
$f'$ stands for the total derivative 
$df/d\xi:=\partial f/\partial \xi\!+\!\hs'\partial f/\partial \hs\!+\!\hD'\partial f/\partial \hD$; $Z$ plays the role of the family parameter. All the other electron dynamical
variables can be expressed in terms of $\hD ,\hs $ and the initial
coordinates $\bX\equiv(X,Y,Z)$ of the generic electron fluid element.
In particular, the dimensionless variable
 $\hbu^{\scriptscriptstyle \perp}:=\hbp^{\scriptscriptstyle \perp}/mc$, i.e.  the electrons' transverse momentum in $mc$ units, in the basic approximation 
is given by  \ $\hbu^{\scriptscriptstyle \perp}=\frac {e\Bap }{mc^2}$; \ hence $v=\hbu^{\scriptscriptstyle \perp}{}^2$. \ The  light-like coordinate 
$\xi=ct\!-\!z$  in Minkowski spacetime can be 
adopted   instead of time $t$ as an independent variable because all particles
must travel at a speed lower than $c$; at the end, to express the solution as a function of $t$ one just needs to replace everywhere $\xi$ by the inverses 
$\tilde\xi(t,Z)$ of the strictly increasing (in $\xi$) functions
$\hat t(\xi,Z):=(\xi\!+\!\hze(\xi,Z))/c$.
  \ Eq. (\ref{heq1}) are Hamilton equations with $\xi,\hD , -\hs $  playing the role of the usual $t,q,p$ and  (dimensionless) Hamiltonian
\bea
\hH( \hD , \hs,\xi;Z)\equiv  \frac{\hs^2+1\!+\!v(\xi)}{2\hs}
+ \U( \hD ;Z);  
%\qquad \U( \Delta;Z)\!\equiv\! K\!\!\int^{\Delta}_0\!\!\!\!\!d\zeta\,(\Delta\!-\!\zeta)
%\,\widetilde{n_0}(Z\!+\! \zeta).
%K\left[\widetilde{{\cal N}}\!\left(Z \!+\!  \Delta\right) \!-\!\widetilde{{\cal N}}\!(Z)\!-\! \widetilde{N}\!(Z) \hD \right],
%=K\!\!\!\!\int\limits^{Z+\Delta}_Z\!\!\!d\zeta\,\widetilde{n_0}(\zeta)  (Z\!+\! \Delta\!-\!\zeta)
%=K\!\int\limits^{z}_Z\!\!d\zeta\,\widetilde{n_0}(\zeta)  (z\!-\!\zeta),  \\[8pt] 
%\quad \displaystyle\widetilde{{\cal N}}(Z)\equiv \int^Z_0\!\!\!d\zeta\,\widetilde{N}(\zeta).
%\!=\!\int^{Z}_0\!\!\!d\zeta\,\widetilde{n_0}(\zeta)  (Z\!-\!\zeta).
                               \label{hamiltonian}
\eea
%(note that it is {\it rational} in $\hs$). 
the first term gives the kinetic $+$ rest mass energy, while $\U$ plays the role of a potential energy due to the
electric charges' mutual interaction.
Consequently, along the solutions of (\ref{heq1}) \ $\hH'=\partial \hH/\partial \xi=v'/2\hs$. \
Integrating  the latter identity by parts, and using the definition (\ref{hamiltonian}) of $\hH$, we find
\bea
\frac{(\hs\!-\!1)^2}{2\hs}+ \U( \hD ;Z)=
 \hH(\xi,Z) - 1%=\int^\xi_0\!\!d\eta\, \frac{v'}{2\hs }(\eta,Z)
-\frac{v(\xi)}{2\hs (\xi,Z)}= \int^\xi_0\!\!d\eta\, \frac{v\hs '}{2\hs ^2}(\eta,Z)=:\nu(\xi,Z).
         \label{Hvar}   
\eea
The Hamilton eqs   (\ref{heq1})  are non-autonomous 
%only
%%
for $0\!<\!\xi\!<\!l$, where  $[0,l]$ is the smallest closed interval containing the support of $\Be^{{\scriptscriptstyle\perp}}$;  ultra-intense pulses are characterized by
$\max_{\xi\in[0,l]}\{v(\xi)\}\gg1$ and induce ultra-relativistic electron motions.
For $\xi\ge l$ (\ref{heq1})  can be solved also by quadrature, using the energy
integrals of motion $\hH(\xi,Z)=\hH(l,Z)=:h(Z)=$ const. 

Solving (\ref{heq1}-\ref{incond})
yields the motions of the $Z$-electrons'  fluid elements, which are fully represented through their worldlines in Minkowski space. 
In fig.  \ref{Worldlines_HomLin-Nonlin}, \ref{Worldlines_Nonlin} we have displayed 
the projections onto the $z,ct$ plane of these worldlines for two specific sets of input data; as we can see, the PW
emerges from them as a collective effect. %of the set of $Z$-electrons motions; 
Mathematically, the PW features can be derived passing to the Eulerian description of the electron fluid; the resulting flow is laminar, with $xy$ plane symmetry.
The Jacobian  of the transformation 
$\bX\mapsto\hbx_e\equiv(\hxe,\hye,\hze)$ 
 from the Lagrangian  to the Eulerian coordinates
reduces to $\hJ(\xi,Z)=\partial\hze(\xi,Z)/\partial Z$, 
because  $\hbx^{\scriptscriptstyle \perp}_e\!-\!\bXp$ does not depend on $\bXp$.

The HD breaks where worldlines intersect, leading to WB of the PW. 
No WB occurs as long as  $\hJ$  remains positive.
If the initial density is uniform, $\widetilde{n_0}(Z)\!=\!n_0\!=\!$ const, 
both the equations  (\ref{heq1}) and the initial conditions (\ref{incond})
become $Z$-independent, because  (\ref{heq1}b)  takes the form  $\hs '=M\hD $, 
where $M\!:=\!Kn_0$. Consequently, also their solutions become $Z$-independent,
and $\hJ\equiv 1$ at all $\xi$. Otherwise, WB   occurs after a sufficiently long time \cite{Daw59}.

Our main goal here is to determine manageable sufficient conditions on $\widetilde{n_0}(z),\Be^{{\scriptscriptstyle\perp}}(\xi)$ guaranteeing that $\hJ(\xi,Z)>0$ 
for all $Z>0$ and $\xi\in[0,l]$,  {\it without} solving the Cauchy problems (\ref{heq1}-\ref{incond}). This will ensure that there is no wave-breaking during the laser-plasma interaction (WBDLPI), i.e. while the Hamilton equations (\ref{heq1}) are non-autonomous (due to the dependence of $v$ on $\xi$). We reach this goal by determining upper and lower bounds first on
$\hD ,\hs ,\hH$ (section \ref{DsHbounds}), then  on $\hJ$ and $\partial \hs /\partial Z$ (section \ref{Jbound}),
also with the help of a suitable Liapunov function. These bounds provide
also useful approximations of these dynamical variables in the interval $0\le\xi\le l$.  
As said, the NR short-pulse conditions (\ref{Lncond}) are generalized by the ones (\ref{Lncond'})
 in the present, fully relativistic regime.
Inequalities (\ref{ShortPulse1}), (\ref{ShortPulse1'}) are respectively sufficient conditions for  (\ref{Lncond'}a), (\ref{Lncond'}b). 
Instead of  (\ref{ShortPulse1}) one may first check the stronger, but also more easily verifiable, condition  (\ref{ShortPulse2}),
or even the simplest (and strongest) one $M_ul^2\le 2$. In case (\ref{ShortPulse1'}) is satisfied, 
we can exclude WBDLPI: in the NR regime, if also (\ref{NR-NoWBcond}) is fulfilled; in the general  case, if one of the three conditions of Proposition 
\ref{PropNoWB} is fulfilled (if $Q_0<1$, the strongest but easiest to compute, is not fulfilled, then one may check $Q_1<1$, or $Q_2<1$), 
namely if initially the plasma is sufficiently diluted and/or the local relative variations of its density
are sufficiently small. In section \ref{discuss} we compare the dynamics of $\hs,\hD,J,\sigma$ induced by the same pulse on five representative $\widetilde{n_0}(z)$ having the same  upper bound and asymptotic value $n_b$, both by  numerically
solving the equations and by applying the mentioned inequalities. In particular, we learn  that  the density  profile
at the very edge of the plasma is critical; for instance, if $\widetilde{n_0}(z)\sim z$ as $z\to 0^+$  then WB occurs earlier
(albeit electrons collide with very small relative velocities)  than if
$\widetilde{n_0}(z)\sim z^2$, or if $\lim_{z\to 0^+}\widetilde{n_0}(z)>0$ (discontinuous density at $z=0$).
To produce LWFA one usually shoots the laser pulse orthogonally to a supersonic gas (e.g. hydrogen or helium) jet; 
since outside the jet nozzle it is $\widetilde{n_0}(z)\sim z^2$, our results imply that
under rather broad conditions such an impact occurs in the hydrodynamic regime.

For $\xi\ge l$, using the conservation of energy, 
one can show \cite{Fio-impact} that, while   $\hD$ and $\hs $
are periodic with a suitable period $\xiH$,  $\hJ$ and $\hsg$ are {\it linearly quasiperiodic}, namely of the form
\be
f(\xi)=a(\xi)+\xi \, b(\xi), \qquad \xi\ge l,
 \label{lin-pseudoper}
\ee
where $a,b$ are periodic in $\xi$ with period $\xiH$, and $b$ has zero average over  a period; 
$b(\xi)$  oscillates between positive and negative values, and so does the second term,
 which dominates as $\xi\to \infty$, with $\xi$ acting as a modulating amplitude.
Therefore the occurrence of WB after the laser-plasma interaction is best
investigated studying the dependence (\ref{lin-pseudoper}) \cite{Fio-impact}. 

 The spacetime region where the present plane hydrodynamic model (predicting a laminar and $xy$-symmetric flow) is self-consistent is determined
(section \ref{discuss}) by the conditions $J>0$ (no collisions) and (\ref{neglectDepletion}) (undepleted pulse approximation); for typical LWFA experiments the length and time sizes allowed by  (\ref{neglectDepletion}) are respectively of the order of several  hundreds of microns, femtoseconds.  The spacetime region where the predictions of the model can be trusted is further reduced (section \ref{discuss})  by  the finite transverse size $R$ of the laser pulse.
According to our model, other phenomena characteristic of plasma physics, like turbulent flows, diffusion, heating, heat exchange, as well  the very motion of ions, can be excluded  inside the latter region, but of course can and will occur outside.

\medskip
Finally, we point out that recent advances in multi-timescale analysis permitted also (semi)analytical studies of laser-plasma interaction  (including some aspects of WB) in the ultrarelativistic regime using  Vlasov description \cite{JovFedBelDeN19}.

\section{Apriori estimates of  $\hD ,\hs ,\hH$ for small $\xi>0$}  
\label{DsHbounds}

\begin{figure}[htbp]
\includegraphics[width=17cm]{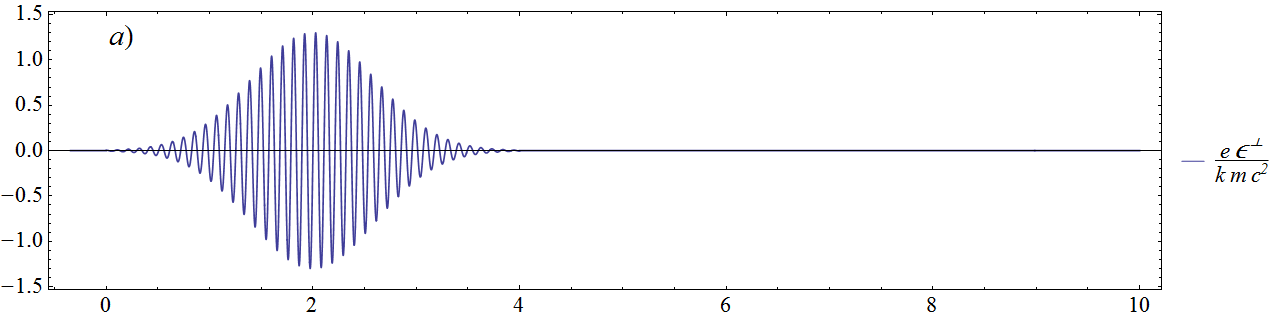}\\
\includegraphics[width=16.55cm]{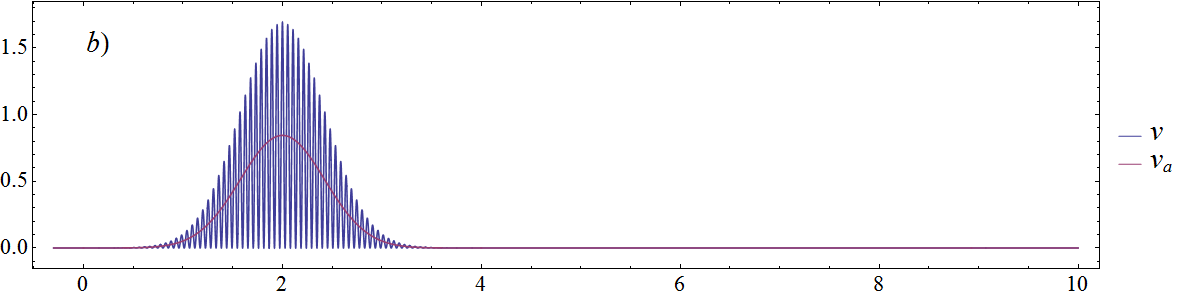} \\
\includegraphics[width=16.88cm]{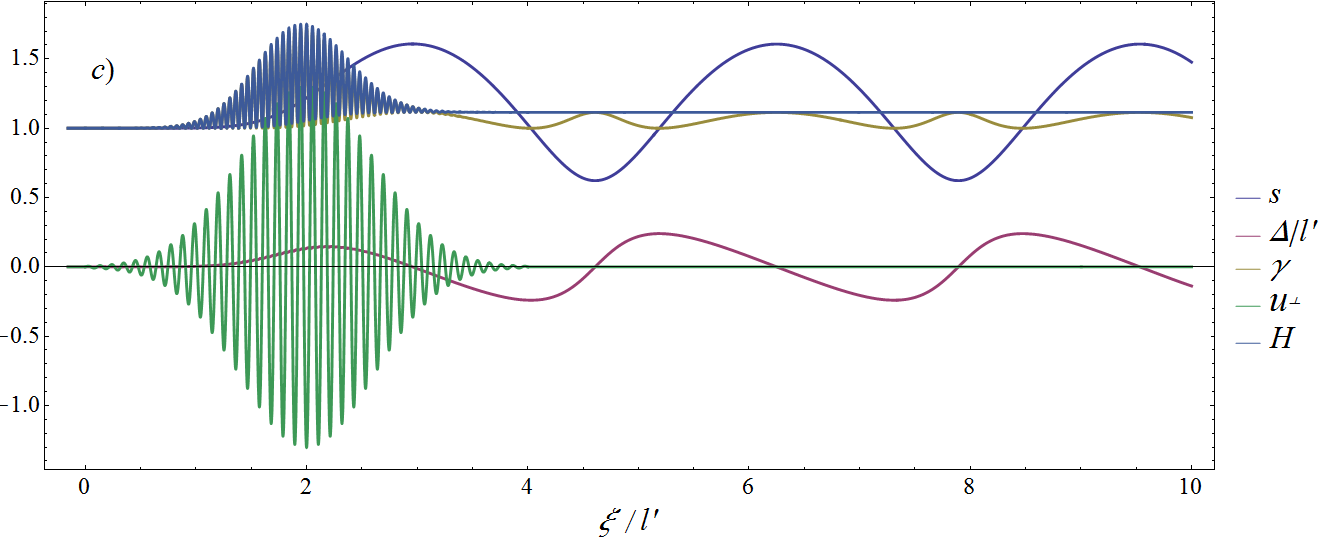} 
\caption{\ a)  Normalized amplitude of a  linearly polarized [i.e. set $\psi=0$ in (\ref{modulate})]
monochromatic laser pulse slowly modulated by a Gaussian with {\it full width at half maximum} $l'$ and 
peak  amplitude $a_0\!\equiv\!\lambda eE^{\scriptscriptstyle \perp}_{\scriptscriptstyle M}/mc^2\!=\!1.3$; this yields a moderately relativistic electron dynamics, and $\Delta_u\equiv\hDO(l)\simeq 0.45 l'$. If $l'\!=\!7.5\mu$m (corresponding to a pulse duration of $\tau'=l'/c\simeq 2.5\times 10^{-14}$s), and the wavelength is $\lambda=0.8\mu$m,
then the corresponding peak  intensity must be $I\!=\!7.25\!\times\!10^{18}$W/cm$^2$;
these are typical values obtainable by Ti:Sapphire lasers in LWFA experiments.
 \ \ b) \ Corresponding forcing term $v(\xi)$, and average-over-a-cycle  (\ref{cycle-average}) $v_a(\xi)$ of the latter.
\ \  c) \ Corresponding solution 
of (\ref{heq1}-\ref{incond}), or equivalently of (\ref{e3a}), if
$Kn_0l'{}^2\simeq 4;%=3.98
$ %$Ml^2=95.44$, 
 this value is obtained if $l'\!=\!7.5\mu$m 
and $\widetilde{n_0}(Z)\!\equiv\!n_0=2\times 10^{18}$cm$^{-3}$ (a typical value of the electron
density used in LWFA experiments). As expected, $s$ is insensitive to the rapid oscillations of $\Bep$ for $\xi\in[0,l]$, while  for $\xi\ge l$ the energy $H$ is conserved, and the solution is periodic.
The length $l$ is determined
on physical grounds; if e.g. the plasma is created locally by the impact  of the 
front of the
pulse itself  on a gas  (e.g. hydrogen or helium), then $[0,l]$ has to contain all points $\xi$  where the pulse intensity is sufficient to ionize the gas. Here for simplicity and conventionally we have fixed it to be $l=4l'$ [the possible inaccuracy of such a cut is very small, because $\epsilon(0^+)=\epsilon(l^-)$ is  $2^{-16}$ times the maximum $\epsilon(l/2)$ of the modulation, i.e. practically zero], what makes $G_b\equiv \sqrt{Kn_0}\,l\simeq 8$. 
}
\label{graphs}
\end{figure}

The Cauchy problem (\ref{heq1}-\ref{incond}) is equivalent to the following integral one:
\be
\hD (\xi,Z)= \int_0^\xi\!\!d\eta\,\frac {1\!+\!v(\eta)}{2\hs ^2(\eta,Z)}-
\frac{\xi}2,\qquad \hs (\xi,Z)-1= \int_0^\xi\!\!\!d\eta\!\!\int^{\hze(\eta,Z)}_Z\!\!\!\!\!\!\!\!\!\!\!\!
dZ' \: K\widetilde{n_0}(Z').
  \label{sDelta}
\ee
If 	$\widetilde{n_0}(Z)\!\equiv\!n_0\!=\!$ const, 
then \ $\U(\Delta)\!=\!M\Delta^2/2$,  $s'=M\Delta$, and (\ref{sDelta}) amount to
\be
s(\xi)=1+\frac M 2\left[-\frac{\xi^2}2\!+\!\!\int^\xi_0\!\!\! d\eta\, (\xi\!-\!\eta)\frac {1\!+\!v}{s^2}(\eta) \right],  \label{e3a}
\ee 
where $M\!:=\!Kn_0$; once  (\ref{e3a}) is solved, 
one obtains $\Delta$ from $\Delta= s'/M$. In fig. \ref{graphs} 
% ?fai i grafici con $\widetilde{n_0}(Z)\!\equiv\!n_0\!=\!3\times 10^{18}$ e $a_0=1.3$
we plot an example of a monochromatic laser pulse  slowly modulated by a Gaussian and 
 the corresponding solution $(s,\Delta)$ %and the associated electron wordlines, 
in a constant density plasma; the qualitative behaviour of the solution remains the same also
if $\widetilde{n_0}(z)\!\neq$const. 
In the NR regime $v\ll 1$, whence  \ 
$|\Delta/l|\ll 1$, $|\delta|\ll 1$, \ where \  $\delta:=\hs\!-\!1$; \ at lowest order
in $\delta,\Delta$ (\ref{heq1}-\ref{incond})
reduce to the equations \ $\delta'=M\Delta$, $\Delta'=v/2-\delta$ \
of a forced NR harmonic oscillator with trivial initial conditions. The solution is
\bea
\Delta(\xi)=\int_0^\xi\!\!d\eta\,\frac {v(\eta)}{2}\,\cos\left[\sqrt{M}(\xi\!-\!\eta)\right],\quad \delta(\xi)=
%\hs(\xi)\!-\!1=
\int_0^\xi\!\!d\eta\,\frac {v(\eta)}{2}\,\sin\left[\sqrt{M}(\xi\!-\!\eta)\right].
   \label{NRn_0solution}
\eea

By (\ref{heq1}b), the zeroes of $\hD(\cdot,\!Z)$ 
are extrema of $\hs(\cdot,\!Z)$, and vice versa, because $\widetilde{N}(Z)$  grows with $Z$.
 Let us recall how $\hD,\hs$ start evolving  from their initial values (\ref{incond}).  
As said, for $\xi\!>\!  0$ all electrons reached by the pulse start to oscillate transversely and
drift forward; % (pushed by the ponderomotive force); 
in fact, $v(\xi )$ \  becomes positive, implying in turn that so does the right-hand side (rhs) of (\ref{heq1}a) and $\hD$; \  
the $Z\!=\!0$ electrons leave behind themselves a  layer of ions %$L_t$ 
of finite thickness 
%$\zeta(t)\!=\!\Delta(t,0)\!=\!\hD[\tilde\xi(t,0),0]$ 
completely evacuated of electrons.
If the density vanished 
 ($\widetilde{n_0}\equiv 0$) then we would obtain
$$
\hs \equiv 1, \qquad \hD (\xi,Z) %\le \int_0^\xi\!\!d\eta\,\frac {v(\eta)}{2\hs ^2(\eta,Z)}
=\int_0^\xi\!\!d\eta\,\frac {v(\eta)}{2}=:\hDO(\xi);
$$
$\hDO(\xi)$ grows with $\xi$ and is almost constant for $\xi>l$ if $v(\xi)\simeq 0$ for  $\xi>l$
(what occurs if the pulse is slowly modulated  (\ref{modulate})).  
On the contrary, since the density is positive, the growth of $\hD$ implies also the growth of
the rhs of (\ref{heq1}b) (because the latter grows with $\hD$), and of $\hs(\xi,\!Z)\!-\!1$.  \ $ \hD(\xi,\!Z)$ keeps growing  as long as $1\!+\!v(\xi)>\hs^2(\xi,\!Z)$, reaches a maximum at $\tilde\xi_1(Z)\equiv$ the smallest  $\xi\!>\!0$ such that the rhs (\ref{heq1}a)  vanishes.
%, and starts decreasing for $\xi>\tilde\xi_1$. in realtà nel caso slowly modulated fa tante piccole oscillazioni con tanti massimi, pur rimanendo positivo
$ \hs(\xi,\!Z)$ keeps growing  as long as $\hD(\xi,\!Z)\!\ge\!0$, reaches a maximum at the first zero $\tilde\xi_2>\tilde\xi_1$ of $ \hD(\xi,\!Z)$ and  decreases  for $\xi>\tilde\xi_2$, while  $\hD(\xi,\!Z)$ is negative. $ \hD(\xi,\!Z)$  reaches a negative minimum at 
$\check\xi_3(Z)\equiv$ the smallest  $\xi\!>\!\tilde\xi_2$ such that the rhs (\ref{heq1}a)  vanishes 
again. 
We also denote by $\tilde\xi_3(Z)$ the smallest  $\xi\!>\!\check\xi_3$ such that $ \hs(\xi,\!Z)=1$, and $\I:=[0,\tilde\xi_3]$. We encourage the reader to single out $\tilde\xi_1,\tilde\xi_2,\check\xi_3,\tilde\xi_3$ for the solution considered in   fig. \ref{graphs} from the graphs of \ref{graphs}c.
If $\Bep$ is slowly modulated  then $v(l)\!\simeq\!0$
(see section \ref{oscill},  appendix 5.4 in \cite{Fio18JPA} for  details); then $\check\xi_3\simeq\tilde\xi_3$ if in addition $l<\check\xi_3$.
%As $\hs'(\xi,Z)$ is positive for all $\xi\in ]0,\tilde\xi_2[$ and negative for all 
%$\xi\in ]\tilde\xi_2,\tilde\xi_3[$, $\hs $ has an isolated maximum in $\tilde\xi_2$.
%Similarly, $\hD$ has a  maximum in $\tilde\xi_1$. For all $Z>0$ we %are going to 
%find  $0<\tilde\xi_1(Z)<\tilde\xi_2(Z)<\tilde\xi_3(Z)<\infty$, $\Delta(\tilde\xi_2,Z)=0$, $\hs (\tilde\xi_3,Z)=1$.

\begin{Definition}
A pulse is  {\it strictly short}, {\it essentially short} w.r.t. $\widetilde{n_0}$ if it respectively fulfills
\bea
 \left\{\!\!\ba{l}\displaystyle \hD(\xi,Z)\ge 0\: ,\\[8pt]  \displaystyle \hs(\xi,Z)\ge 1\: ,\ea\right. \quad \forall \xi\in[0,l],\: Z\ge 0\qquad \Leftrightarrow \qquad 
l\: \le \: \left\{\!\!\ba{l} \tilde\xi_2(Z)\\[4pt] \tilde\xi_3(Z)\ea\right.  \quad   \forall  Z\ge 0\,.
   \label{Lncond'}
\eea
\label{Def1}
\end{Definition}
In the NR regime if $\widetilde{n_0}(Z)\!\equiv\!n_0\!= $const these respectively amount
to requiring that the corresponding solution (\ref{NRn_0solution}) fulfills $\Delta(l)\ge 0$,
 $\delta(l)\ge 0$; if $\widetilde{n_0}(Z)\!\!\neq$const it is sufficient to replace $n_0$ by $n_b$
to obtain sufficient conditions for the fulfillment of (\ref{Lncond'}); they are 
\bea
 \left\{\!\!\ba{l}\displaystyle 2\Delta(\xi)=\int_0^\xi\!\!d\eta\,v(\eta)\,\cos\left[\sqrt{Kn_b}(\xi\!-\!\eta)\right]\ge 0\: ,\\[8pt]  \displaystyle 2\delta(\xi)=\int_0^\xi\!\!d\eta\,v(\eta)\,\sin\left[\sqrt{Kn_b}(\xi\!-\!\eta)\right]\ge 0\: ,\ea\right. \quad \forall \xi\in[0,l] .    
   \label{LncondNR}
\eea
\begin{prop}
If $v$ is symmetric about $\xi=l/2$, i.e. $v(\xi)=v(l\!-\!\xi)$, then (\ref{LncondNR}a) amounts to $G_b\le 1/2$;
if $v$ in addition has a unique maximum in $\xi=l/2$, then (\ref{LncondNR}b) amounts to $G_b\le 1$.
\label{propshort}
\end{prop}
 The proof is in the appendix (section \ref{Proofpropshort}). \ The assumption that $v(\xi)$ be symmetric is satisfied with very good approximation if the pulse
is a slowly modulated one (\ref{modulate}) with a symmetric modulation $\epsilon(\xi)$ about $\xi=l/2$  (as in fig. \ref{graphs}b), by  (\ref{slowmodappr}).

\medskip
The following estimates hold for $\xi\in\I=[0,\tilde\xi_3]$.
First,   $\hs \ge 1$ and (\ref{sDelta}a)  imply  the  bound
\be
\hD (\xi,Z) \le %\int_0^\xi\!\!d\eta\,\frac {v(\eta)}{2\hs ^2(\eta,Z)}
%\le \int_0^\xi\!\!d\eta\,\frac {v(\eta)}{2}=:
\hDO(\xi).   \label{Deltabound0}
\ee
%$\hDO(\xi)$ is the displacement for zero density and grows with $\xi$; we denote 
%$\DM^{{\!\scriptscriptstyle (0)}}:=\hDO(l)$.

\subsection{Constant density case}

If 	$\widetilde{n_0}(Z)\!\equiv\!n_0\!>\!0$,
for  $\xi\in\I$ we find  by (\ref{Deltabound0}) \
$s(\xi)\!-\!1= M\int_0^\xi\!\!d\eta\, \Delta(\eta)
\le M\int_0^\xi\!\!d\eta\,\hDO(\eta)$, \ i.e.
\be
s(\xi)\le 1 +\frac {M}{2}\int_0^\xi\!\!d\eta\,(\xi-\eta)\,v(\eta)=:\sU(\xi). \label{sbound1-n_0}
\ee
$\sU(\xi)$ grows strictly with $\xi$ and is convex.
Eq. (\ref{sbound1-n_0})  and (\ref{sDelta}a) in turn imply
\bea
\Delta(\xi)
%&\ge& \int_0^\xi\!\!d\eta\,\frac {1\!+\!v(\eta)}{2\left[1\!+\!K \nM\int_0^\eta\!\!d\zeta\,\hD (\zeta,Z)\right]^2}-\frac{\xi}2\\
&\ge& \int_0^\xi\!\!d\eta\,\frac {1\!+\!v(\eta)}{2\left[\sU(\eta)\right]^2}-
\frac{\xi}2=:\hDU(\xi).
 \label{Deltabound1'-n_0}
\eea
$\hDU(\xi)$ vanishes at $\xi=0$, grows with $\xi$ for small $\xi>0$ until its (unique) maximum point;  for larger $\xi$ it decreases and becomes negative. Hence, 
a lower bound  $\tilde\xi_2^{{\scriptscriptstyle (1)}}$
for $\tilde\xi_2$ is the smallest $\xi>0$  such that \ $\hDU(\xi)=0$.  Therefore 
$\hDU(l)\ge 0$  ensures that $\tilde\xi_2\ge\tilde\xi_2^{{\scriptscriptstyle (1)}}\ge l$. 

Eq. (\ref{sDelta}b) also  implies \
$s(\xi)\!-\!1= M\int_0^\xi\!\!d\eta\, \Delta(\eta)\ge M\! \int_0^\xi\!\!d\eta\,\hDU(\eta)$, \ namely
\bea
s(\xi)\ge 1+ 
\frac{M}2\, f(\xi)=:\sD(\xi),\qquad
 f(\xi):=\int_0^\xi\!\!d\eta\,\frac {(\xi\!-\!\eta)[1\!+\!v(\eta)]}{\left[\sU(\eta)\right]^2}- \frac{\xi^2}2. \label{sbound1'0-n_0}
\eea
At least for small $\xi$, this is a more stringent lower bound for $s$ in $\I$ than $ s\ge 1$: \
$f(\xi)$ vanishes at $\xi=0$, grows with $\xi$ for small $\xi>0$ until its (unique) maximum point;  for larger $\xi$ it decreases and becomes negative. 
Hence, a lower bound  $\tilde\xi_3^{{\scriptscriptstyle (1)}}$
for $\tilde\xi_3$ is the smallest $\xi>0$  such that \ $f(\xi)=0$. \ Therefore 
$f(l)\ge 0$  ensures that $\tilde\xi_3\ge\tilde\xi_3^{{\scriptscriptstyle (1)}}\ge l$.

\subsection{Generic density case}

Let $\check n(\xi,Z)\!:=\!\widetilde{n_0}\big[\hze(\xi,Z)\big]$,
$\tilde\xi_2':=\min\{\tilde\xi_2,l\}$,  
$n_u,n_d>0$ be some upper, lower bounds on  $\check n$
\be
n_d(Z)\le \check n(\xi,Z)\le n_u(Z)         \label{n-bounds0}
\ee
for $0\le\xi\le\tilde\xi_2'$. If 	$\widetilde{n_0}(Z)\!\equiv\!n_0$ then $\check n\!\equiv\!n_0$,
and we can set $n_u=n_d=n_0$. In general, a $Z$-independent choice of $n_u$ is
$n_u=n_b$, see (\ref{n_0bounds}).
More accurately, given $\hDOu>0$ such that
\be
\hD (\xi,Z) \le \hDOu \qquad \forall \xi\in[0,\tilde\xi_3],
\ee
then $0\le\hD (\xi,Z) \le \hDOu(Z)$ for all $\xi\in[0,\tilde\xi_2]$
(by the definition of $\tilde\xi_2$), and (\ref{n-bounds0}) holds choosing 
 %$ n_u\!:=\!\sup\limits_{Z\ge 0}\{\widetilde{n_0}(Z)\}$, with $n_d\!\equiv\!\inf\limits_{Z\ge 0}\{\widetilde{n_0}(Z)\}\!\ge\!0$
\be
n_u(Z)\!=\!\max\limits_{Z'\in\big[Z,Z+\hDOu\big]}\{\widetilde{n_0}(Z')\},%=:n_u^1(Z), 
\qquad n_d(Z)\!=\!\min\limits_{Z'\in\big[Z,Z+\hDOu\big]}\{\widetilde{n_0}(Z')\};%=:n_d^1(Z),
\label{n-bounds}
\ee
in general, (\ref{n-bounds}a) is a lower (and therefore better) upper bound  than $n_u=n_b$.
We also abbreviate \ $M_u(Z):=Kn_u(Z)$, \ $M_d(Z):=Kn_d(Z)$. \ 
By  (\ref{Deltabound0}), we can adopt the simple choice 
 $\hDOu:=\hDO(l)$\footnote{Actually, if $v(l)\simeq 0$, as it occurs if the pulse is a slowly modulated one (\ref{modulate}),
then $\hDO(\xi)\simeq \hDO(l)$ if $\xi>l$, and (\ref{n-bounds0}) holds 
 with  $\hDOu=\hDO(l)$ for all $0\le\xi\le\tilde\xi_2$, even if $\tilde\xi_2>l$.}. 

\begin{lemma} For all $\xi\in[0,\tilde\xi_3]$  the rhs $\nu$ of (\ref{Hvar}) can be bound by
\bea
%\frac{M_d(Z)}{2}\left[\hDU(\xi,Z)\right]^2\le
\nu(\xi,Z)%\int^\xi_0\!\!d\eta\, \frac{v\hs '}{2\hs ^2}(\eta,Z) 
\le  \nu_u(\xi,Z) :=\frac{M_u(Z)}{2}\left[\hDO(\xi)\right]^2
   \label{lemma1} 
\eea
\end{lemma}
\bp{} For $\xi\in[0,\tilde\xi_2]$ the inequality is proved as follows:
\bea
\nu(\xi)=\!\! \int^\xi_0\!\!\!\!d\eta\, \frac{v\hs '}{2\hs ^2}(\eta)\le \!\!\int^\xi_0\!\!\!\!d\eta\, \frac{v }{2\hs ^2}(\eta)\!\!\int^{Z+\hD(\eta)}_Z\!\!\!\!\!\!\!\!\!\!\!\!\!\!dz\,M_u 
= M_u\!\! \int^\xi_0\!\!\!\!d\eta\, \frac{v\hD }{2\hs ^2}(\eta)
%\le M_u \int^\xi_0\!\!d\eta\, \frac{v\hDD}{2\hsD{}^2}(\eta)
\le M_u\!\! \int^\xi_0\!\!\!\!d\eta\, \frac{v\hDO}{2}(\eta)= \nu_u(\xi)
% \frac{M_u}{2}\left[\hDO(\xi)\right]^2,
\nonumber
\eea
where we have used $\hDO{}'=v/2$ and  for brevity we have not displayed the $Z$ argument. 
The inequality holds also  in $]\tilde\xi_2,\tilde\xi_3]$ because there $\nu$ decreases, whereas $\nu_u$ grows.
\ep

 The maximum of $\nu(\xi,Z)$ in  $[0,\tilde\xi_3]$  is in $\xi=\tilde\xi_2$ 
 because $s'> 0$ in $]0,\tilde\xi_2[$ and  $s'<0$ in $]\tilde\xi_2,\tilde\xi_3[$.
To obtain  upper, lower bounds $\hs_u,\hs_d$ for $\hs(\xi,Z)$ and a lower bound
$\hD_d(Z)$ for $\hD(\xi,Z)$ in the longer interval $0\le\xi\le\tilde\xi_3':=\min\{l,\tilde\xi_3\}$\footnote{Again, 
if $v(l)\simeq 0$, as it occurs if the pulse is a slowly modulated one (\ref{modulate}),
then $\hDO(\xi)\simeq \hDO(l)$ if $\xi>l$, and these results remain valid 
if we replace $\tilde\xi_3'$ by  $\tilde\xi_3$, even if $\tilde\xi_3>l$.} 
we use (\ref{lemma1}) to majorize 
$$
\nu(\xi,Z)\le \nu\big(\tilde\xi_2,Z\big) \le   \frac{M_u}{2}\left[\hDO\big(\tilde\xi_2\big)\right]^2\le  \frac{M_u}{2}\hD_u^2.
$$
%Nota che con questa dimostrazione, non posso automaticamente sostituire $\hDO(l)$ con qualche
%altro $\Delta_u$ più stringente
This, replaced in (\ref{Hvar}), yields \ $(\hs\!-\!1)^2/2\hs\le M_u\hD_u^2/2$ \ and \
$\U\le  M_u\hD_u^2/2$, \ whence
\be
% \U\le   \frac{M_u}{2}\hD_u^2 \frac{(\hs\!-\!1)^2}{2\hs}\le\frac{M_u}{2}\hD_u^2
%\quad\Rightarrow\quad 
\hs_d\le\hs(\xi,Z)\le \hs_u, \:\: 
\left.\ba{l}\hs_u\\ \hs_d\ea\!\!\right\}
:=1\!+\!\frac{M_u}{2}\hD_u^2\!\pm\!\sqrt{\!\left(\!1\!+\!\frac{M_u}{2}\hD_u^2\!\right)^2\!-\!1} ,\qquad 
\Delta(\xi,Z)\ge \hDd(Z),   \label{sDeltabound'}
\ee
where $\hDd$ is the negative solution of the equation $ \U( \hD ;Z)= M_u(Z)\hD_u^2/2$
 \ (as a first estimate, $\hDd=-\Delta_u$). Clearly, \ $1/\hs_d=\hs_u>1$.
%\be
%  \U( \hD ;Z)\le  \frac{M_u}{2}\left[\hDO(l)\right]^2.\ee

\begin{prop}
 For all $\xi\!\in\![0,\tilde\xi_3']$ the dynamical variables $\hD ,\hs $ are bounded as follows:
\bea
\ba{l}
 \hDO(\xi,Z)\ge\hD (\xi,Z) \ge \hDU(\xi,Z), \\[12pt]
 \hsU(\xi,Z)\ge\hs (\xi,Z) \ge \hsD(\xi,Z),
\ea
\label{sDeltabound1}
\eea
where  \ %for $\xi\in[0,\tilde\xi_2]$
$\hDO(\xi)\!:=\!\! \int\limits_0^\xi\!  d\eta\, v(\eta)/2$, \ and
\bea
\ba{l}
\displaystyle 
% \hDO(\xi)\!:=\!\! \int_0^\xi\!\!\!\! d\eta\,\frac {v(\eta)}{2}, \qquad
\hsU(\xi,\!Z):=\min\left\{\hs_u,1 +g(\xi,\!Z)\right\},\quad
g(\xi,\!Z):=\frac {M_u}{2}\!\!\int_0^\xi\!\!\!\!d\eta\,(\xi\!-\!\eta)\,v(\eta),  \\[12pt]
\displaystyle  \hDU(\xi,\!Z):=\max\left\{\hDd,d(\xi,\!Z)\right\},\quad d(\xi,\!Z):=
\!\!\int_0^\xi\!\!\!\!d\eta\,\frac {1\!+\!v(\eta)}{2\left[\hsU(\eta,\!Z)\right]^2}-
\frac{\xi}2 ,\\[14pt]
\displaystyle  
%\hDO(\xi)\!:=\!\!  \int_0^\xi\!\!\!d\eta\,\frac {v(\eta)}{2}, \qquad
\hsD(\xi,Z):= \left\{\!\! \ba{ll}
1+ \frac{M_d}2  f(\xi,Z) \quad & 0\le\xi\le \tilde\xi_2^{{\scriptscriptstyle (1)}}\\[8pt]
\max\left\{\hs_d,1+ \left[\frac{M_d}2\!-\! \frac{M_u'}2\right]  f\left(\tilde\xi_2^{{\scriptscriptstyle (1)}},Z\right) +
\frac{M_u'}2\,f(\xi,Z)\right\}\quad & \tilde\xi_2^{{\scriptscriptstyle (1)}}< \xi\le \tilde\xi_3',\ea\right. \\[20pt]
\displaystyle f(\xi,Z):=\!\!\int_0^\xi\!\!\! d\eta\,(\xi\!-\!\eta)\!\left\{\!\frac {1\!+\!v(\eta)}{\left[\hsU(\eta,Z)\right]^2}- 1\!\right\}, 
\ea          \label{sDelta1Def}       
\eea
where $\tilde\xi_2^{{\scriptscriptstyle (1)}}(Z)<\tilde\xi_2$ is the zero of $d(\xi,\!Z)$ as well as the
maximum point of $ f(\xi,Z)$, and $M_u'/K=n_u'\!:=\!\max\limits_{Z'\in\left[Z+\hDd,Z\right]}\{\widetilde{n_0}(Z')\}$.
Moreover, the value of the Hamiltonian  is bounded by
\bea
\Hd:=1+\frac{v}{2\hsU}+\nu_d\:\le\: H \:\le\: 1+\frac{v}{2\hsD} + \frac{M_u}{2}\left[\hDO\right]^2=: \Hu, 
 \label{Hbound}
\eea
where, dubbing the maximum of $d(\xi,\!Z)$  by \ $\tilde\xi_1^{{\scriptscriptstyle (1)}}(Z)$, we have defined
\be
 \nu_d(\xi,Z) :=\left\{\!\!\ba{ll} \displaystyle\frac{M_d(Z)}{2}\left[\hDU(\xi,Z)\right]^2\quad &\xi\in\big[0,\tilde\xi_1^{{\scriptscriptstyle (1)}}\big],\\[8pt]
\displaystyle  \frac{M_d(Z)}{2}\left[\hDU\big(\tilde\xi_1^{{\scriptscriptstyle (1)}}\big)\right]^2\quad &\xi\in\big]\tilde\xi_1^{{\scriptscriptstyle (1)}},\tilde\xi_2^{{\scriptscriptstyle (1)}}\big],\\[8pt]
\displaystyle \frac{M_d(Z)}{2}\left[\hDU\big(\tilde\xi_1^{{\scriptscriptstyle (1)}},Z\big)\right]^2\!+M_u'(Z)\!  \int_{\tilde\xi_2^{{\scriptscriptstyle (1)}}}^\xi\!\! \! d\eta\,\frac{v\hDU}{2} (\eta,Z)\quad &\xi\in\big]\tilde\xi_2^{{\scriptscriptstyle (1)}},\tilde\xi_3'\big].\ea\right.
  \label{Defnu_d} 
\ee
\label{PropBounds-sDeltaH}
\end{prop}
[Note that $\tilde\xi_1^{{\scriptscriptstyle (1)}}(Z)<\tilde\xi_1(Z)$]. \
Eq. (\ref{sDeltabound1}-\ref{sDelta1Def}) reduce to 
(\ref{Deltabound0}-\ref{sbound1'0-n_0}) if \ 	$\widetilde{n_0}(Z) \equiv n_0=$ const.

\bp{}
The left inequality in (\ref{sDeltabound1}a) is the already proven  (\ref{Deltabound0}).
Eq. (\ref{sDelta}b) by (\ref{n-bounds}),  (\ref{Deltabound0})  implies \ 
$\hs (\xi,Z)- 1\le\int_0^\xi\!\!d\eta\,M_u\hD (\eta,Z)
\le M_u\int_0^\xi\!\!d\eta\,\hDO(\eta)$, \ what, together with (\ref{sDeltabound'}a), implies 
the left inequality in (\ref{sDeltabound1}b);
the latter, together with (\ref{sDelta}a), (\ref{sDeltabound'}b)  in turn imply the right inequality in (\ref{sDeltabound1}a). $d(\xi,Z)=f'(\xi,Z)$ vanishes at $\xi=0$, grows with $\xi$ for small $\xi>0$,
as far as it reaches a maximum
 for sufficiently large $\xi$;  then 
decreases to negative values. Hence
$\tilde\xi_2^{{\scriptscriptstyle (1)}}$, i.e. the smallest $\xi>0$ such that $\hDU(\xi,Z)=0$, 
 is indeed  a lower bound for $\tilde\xi_2$; 
\ $\tilde\xi_2^{{\scriptscriptstyle (1)}}$ is also the maximum point of $f$ and  $\hsD$. 
Eq. (\ref{sDelta}b) also implies for all $\xi\in[0,\tilde\xi_2]$
\be
\hs (\xi,\!Z)\!-\!1\ge M_d\! \int_0^\xi\!\!d\eta\,\hD (\eta,\!Z)\ge M_d\! \int_0^\xi\!\!d\eta\,d(\eta,\!Z)=\frac{M_d}2  f(\xi,Z).      \label{proofstep1} 
\ee
If $\xi\!\in]\tilde\xi_2,\tilde\xi_3']$, integrating (\ref{heq1}b) over $]\tilde\xi_2,\xi]$ and recalling that $\hDd\le\hD< 0$ there, we find
\bea
\hs (\xi,\!Z)\!-\!\hs (\tilde\xi_2,\!Z) &=& -K\! \int_{\tilde\xi_2}^\xi\!\!\!\!d\eta\,
 \int_{Z+\hD (\eta,\!Z)}^Z\!\!\!\!\!\!dZ'\,\widetilde{n_0}(Z')
\ge M_u'\! \int_{\tilde\xi_2}^\xi\!\!d\eta\,\hD (\eta,\!Z)\ge M_u'\! \int_{\tilde\xi_2}^\xi\!\!d\eta\,\hDU (\eta,\!Z) \nn
& > & M_u'\! \int_{\tilde\xi_2^{{\scriptscriptstyle (1)}}}^\xi\!\!d\eta\,\hDU (\eta,\!Z)
\ge \frac{M_u'}2\,\left[f(\xi,Z)-f\left(\tilde\xi_2^{{\scriptscriptstyle (1)}},Z\right)\right]
 \qquad  \label{proofstep2} 
%\nonumber
\eea
(the first inequality in the last line holds because $\hDU (\eta,\!Z)<0$ if 
$\eta\in]\tilde\xi_2^{{\scriptscriptstyle (1)}},\tilde\xi_2]$); since $\hs$ has its maximum in $\tilde \xi_2$, (\ref{proofstep1}) implies in particular $\hs \big(\tilde\xi_2,\!Z\big)\ge \hs \big(\tilde\xi_2^{{\scriptscriptstyle (1)}},\!Z\big)\ge1\!+\! \frac{M_d}2  f \big(\tilde\xi_2^{{\scriptscriptstyle (1)}},\!Z\big)$, which replaced in  
 (\ref{proofstep2})  gives
$$
\hs (\xi,\!Z)\ge 1\!+\! \frac{M_d}2  f \big(\tilde\xi_2^{{\scriptscriptstyle (1)}},\!Z\big)+
 \frac{M_u'}2\,\left[f(\xi,Z)-f\left(\tilde\xi_2^{{\scriptscriptstyle (1)}},Z\right)\right];
$$
the latter inequality and (\ref{proofstep1}) amount to the right inequality in
(\ref{sDeltabound1}b), which holds together with $\hs (\xi,Z) \ge 1$.
The right inequality in (\ref{Hbound}) follows from  (\ref{Hvar}) and (\ref{lemma1}).
From \ $2\hDU{}'=(1\!+\!v)/\hsU{}^2\!-\!1\le v/\hsU{}^2$ \ it follows for $\xi\in[0,\tilde\xi_2]$
\bea
\nu(\xi,Z) %\int^\xi_0\!\!d\eta\, \frac{v\hs '}{2\hs ^2}(\eta) 
\ge M_d \!\!\int^\xi_0\!\!\!\!d\eta\, \frac{v\hD }{2\hs ^2}(\eta)
%\ge M_u\!\! \int^\xi_0\!\!d\eta\, \frac{v\hDD}{2\hsD{}^2}(\eta)
\ge M_d\!\! \int^\xi_0\!\!\!\!d\eta\, \frac{v\hDU}{2\hsU{}^2}(\eta)\ge  M_d \!\!\int^\xi_0\!\!\!\!d\eta \left[\hDU\hDU{}'\right]\!(\eta)=  \frac{M_d}{2}\left[\hDU(\xi)\right]^2,
\nonumber
\eea
where for brevity again we have not displayed the $Z$ argument; since the rhs has its maximum
in $\xi=\tilde\xi_1^{{\scriptscriptstyle (1)}}$, whereas $\nu(\xi,Z)$ still grows in 
$]\tilde\xi_1^{{\scriptscriptstyle (1)}},\tilde\xi_2\big]$, we obtain
\be
\nu(\xi,Z) \ge\left\{\!\!\ba{ll} \frac{M_d}{2}\left[\hDU(\xi)\right]^2\quad &\xi\in\big[0,\tilde\xi_1^{{\scriptscriptstyle (1)}}\big]\\[4pt]
 \frac{M_d}{2}\left[\hDU\big(\tilde\xi_1^{{\scriptscriptstyle (1)}}\big)\right]^2\quad &\xi\in\big]\tilde\xi_1^{{\scriptscriptstyle (1)}},\tilde\xi_2\big]\ea\right.
  \label{proofstep3} 
\ee
If $\xi\in]\tilde\xi_2,\tilde\xi_3']$ then $\hs',\hD<0$ in $ ]\tilde\xi_2,\xi]$, and 
\bea
\int^\xi_{\tilde\xi_2}\!\!\! \!d\eta\, \frac{v\hs '}{2\hs ^2}(\eta) =-\!\! \int_{\tilde\xi_2}^\xi\!\!\!\!d\eta\,
\frac{Kv }{2\hs ^2}(\eta) \!\!\!\int_{Z+\hD (\eta)}^Z\!\!\!\!\!\!\!\!\!\! \!\!dz\,\widetilde{n_0}(z)
\ge M_u'\! \int_{\tilde\xi_2}^\xi\!\! \!\! d\eta\,\frac{v\hD}{2}(\eta)\ge M_u' \!\! \int_{\tilde\xi_2}^\xi\!\! \!\!d\eta\,\frac{v\hDU}{2} (\eta)> M_u' \!\! \int_{\tilde\xi_2^{{\scriptscriptstyle (1)}}}^\xi\!\! \!\!d\eta\,\frac{v\hDU}{2} (\eta)
%\qquad\Rightarrow\qquad
\nonumber
\eea
(the last inequality holds because $\hDU<0$ in $\big]\tilde\xi_2^{{\scriptscriptstyle (1)}},\tilde\xi_2\big]$); summing this inequality and \ $\nu\big(\tilde\xi_2,Z\big)\ge M_d\left[\hDU\big(\tilde\xi_1^{{\scriptscriptstyle (1)}}\big)\right]^2\!/2$ \ we obtain  the one
\ $\nu\big(\xi,Z\big)\ge M_d\left[\hDU\big(\tilde\xi_1^{{\scriptscriptstyle (1)}}\big)\right]^2\!/2
\!+\!M_u' \!\!\displaystyle \int_{\tilde\xi_2^{{\scriptscriptstyle (1)}}}^\xi\!\! d\eta\,v\hDU(\eta)/2$, \ which actually holds  for all $\xi\!\in]\tilde\xi_2^{{\scriptscriptstyle (1)}},\tilde\xi_3']$, by
(\ref{proofstep3}) and
because the second term is negative if  $\xi\!\in]\tilde\xi_2^{{\scriptscriptstyle (1)}},\tilde\xi_2]$.
Summing up, we find $\nu(\xi,Z) \ge \nu_d(\xi,Z)$ in all the interval $[0,\tilde\xi_3']$; this, together
with  (\ref{Hvar}), implies also the left inequality (\ref{Hbound}).
\ep

\noindent
By (\ref{sDeltabound'}), inequalities (\ref{n-bounds0})  hold for all $\xi\in [0,\tilde\xi_3']$ if, rather than by (\ref{n-bounds}),
 we define $n_u,n_d$ by
\be
n_u(Z)\!=\!%n_u^2(Z)\!:=\!
\max\limits_{Z'\in\big[Z+\hDd,Z+\hDOu\big]}\{\widetilde{n_0}(Z')\}, \qquad n_d(Z)\!=\!
%n_d^2(Z)\!:=\!
\min\limits_{Z'\in\big[Z+\hDd,Z+\hDOu\big]}\{\widetilde{n_0}(Z')\}.
\label{n-bounds'}
\ee

As said, $\hDU(\xi,Z)=d(\xi,Z)=f'(\xi,Z)$ vanishes at $\xi=0$, 
grows up to its unique positive  maximum at $\tilde\xi_1^{{\scriptscriptstyle (1)}}$,
then decreases to negative values; $\tilde\xi_2^{{\scriptscriptstyle (1)}}$ is the unique 
$\xi>\tilde\xi_1^{{\scriptscriptstyle (1)}}$ such that $\hDU(\xi,Z)=0$. Hence, 
$\tilde\xi_2^{{\scriptscriptstyle (1)}}$ is a lower bound for 
 $\tilde\xi_2$. Therefore the condition 
\be
\hDU(l,Z)\ge 0                                   \label{ShortPulse1}
\ee 
ensures that $\tilde\xi_2(Z)\ge\tilde\xi_2^{{\scriptscriptstyle (1)}}(Z)\ge l$, namely that
the pulse is strictly short. Similarly,  $\hsD\!-\!1$
vanishes at $\xi=0$, 
grows up to its unique positive positive maximum at $\tilde\xi_2^{{\scriptscriptstyle (1)}}$,
then decreases to negative values. Hence, 
a lower bound $\tilde\xi_3^{{\scriptscriptstyle (1)}}$ for  $\tilde\xi_3$   is the unique 
$\xi>\tilde\xi_2^{{\scriptscriptstyle (1)}}$ such that $\hsD(\xi,Z)=1$, and the condition 
\be
\hsD(l,Z)\ge 1                                   \label{ShortPulse1'}
\ee 
ensures that $\tilde\xi_3(Z)\ge\tilde\xi_3^{{\scriptscriptstyle (1)}}(Z)\ge l\equiv \tilde\xi_3'$, namely that
the pulse is essentially short.
%(implying that the longitudinal displacement of the the $Z$-electrons has not reached its negative minimum yet). QUASI

Moreover, under this assumption we find
%: i) $\hs (\tilde\xi_2,Z) \ge \hs \big(\tilde\xi_2^{{\scriptscriptstyle (1)}},Z\big) \ge \hsD\big(\tilde\xi_2^{{\scriptscriptstyle (1)}},Z\big)$; ii)   
by (\ref{Hbound})
the following upper and lower bounds on the final $Z$-electron energy $h(Z)$ after
their interaction with the pulse:
\bea
\Hd(l,Z) \:\le\: h(Z) \:\le\: \Hu(l,Z).   \label{hbound}
\eea
In fig. \ref{graphs-bounds_Homlin-LinInhom-const-1}, \ref{graphs-bounds_Homlin-Nonlin} we  plot $\hs(\cdot,Z),\Delta(\cdot,Z)$ for  two values of $Z$ and the associated upper and lower bounds corresponding to the densities of fig. \ref{InDensityPlots},  which have the same asymptotic value $n_b$.
As we can see, the bounds agree well. 
Moreover, a useful lower bound for $\hDU$ is  provided by 

\begin{lemma}
.\vskip-1.5cm
\bea
\qquad \qquad \qquad 
\hDU(\xi,\!Z)\ge  \hDO(\xi)\left[1-\frac{M_u(Z)}2  \xi^2 -M_u(Z)\,  \xi \,\hDO(\xi)\right].
\label{EasyBound-Delta1}
\eea
\end{lemma}
\bp{}
$g\ge 0$ implies $1/(1\!+\!g)\ge 1\!-\!g$, $1/(1\!+\!g)^2\ge (1\!-\!g)^2\ge 1\!-\!2g$, so that
\bea
2\hDU(\xi,\!Z)\ge\!\!\int_0^\xi\!\!\!\!d\eta\,\left\{[1\!+\!v(\eta)] \left[ 1\!-\!2g(\eta,Z)\right]-
1\right\}=2\hDO(\xi)-\!\!\int_0^\xi\!\!\!\!d\eta\,2g(\eta,Z)[1\!+\!v(\eta)].
\label{temp-Delta1}
\eea
The definitions of $g,\hDO$ immediately imply the inequality \ $g(\xi,\!Z)\le M_u  \xi \hDO(\xi)$, \ whence
\bea
%g(\xi,\!Z)\le M_u  \xi \hDO(\xi),\quad 
\int_0^\xi\!\!\!\!d\eta\,2g(\eta,\!Z)\le M_u\!\!   \int_0^\xi\!\!\!\!d\eta\,\eta 2\hDO(\eta)
\le M_u\!\!   \int_0^\xi\!\!\!\!d\eta\,\eta 2\hDO(\xi)\le  M_u   \xi^2 \hDO(\xi),\quad \nn
 \int_0^\xi\!\!\!\!d\eta\,2g(\eta,\!Z) v(\eta)\le M_u\!\!   \int_0^\xi\!\!\!\!d\eta\,\eta 2\hDO(\eta)v(\eta)\le M_u\!\!   \int_0^\xi\!\!\!\!d\eta\,\xi 4\left[\hDO\hDO{}'\right](\eta)=
 2 M_u   \xi [\hDO(\xi)]^2\nonumber
\eea
Replacing the latter in (\ref{temp-Delta1}) we obtain (\ref{EasyBound-Delta1}). \ep

As a consequence, if the square bracket at the rhs(\ref{EasyBound-Delta1}) is nonnegative, then 
so is $\hDU(\xi,\!Z)$, and therefore another  condition ensuring that $\tilde\xi_2> l$
(i.e. that the pulse is strictly short) is
\bea
M_u(Z)\,  l^2\left[1 + 2\frac{\Delta_u}l\right] \le 2,         \label{ShortPulse2}
\eea
which is  more easily computable, but also more difficult to satisfy, than (\ref{ShortPulse1}).

\bigskip
One could find more stringent (but also less easily computable) 
bounds than (\ref{sDeltabound1})  by replacing them 
in (\ref{sDelta}) and reiterating the previous arguments\footnote{The first step is
the new bound  $\hD (\xi,Z)\le \hDD(\xi,Z):=\int_0^\xi\!\!d\eta\,\frac {1\!+\!v(\eta)}{2\left[\hsD(\eta,Z)\right]^2}-
\frac{\xi}2 $, the second is setting $\hDOu:=\max\limits_{\xi\in[0,l]}\{\hDD(\xi)\}$ instead of $\hDOu:=\hDO(l)$ in (\ref{n-bounds}), and so on.}, but for the scope of the present work we content ourselves with these basic relations.

\section{Bounds on the Jacobian for small $\xi>0$}
\label{Jbound}

Differentiating (\ref{heq1}) we find that the dimensionless variables
\be
\varepsilon(\xi,Z):=\hJ(\xi,Z)\!-\!1=\frac{\partial\hs(\xi,Z)}{\partial Z},\qquad
\sigma(\xi,Z):=l\frac{\partial\hs(\xi,Z)}{\partial Z}
\ee
 fulfill the Cauchy problem
\bea
\ba{ll}
\varepsilon'=-\kappa\sigma, \qquad 
&\sigma'=Kl\left(
\check n -\! \widetilde{n_0}+\check n\,\varepsilon\right), \\[10pt]
\varepsilon(0,Z)=0,\qquad &\sigma(0,Z)=0,  
\ea          \label{basic1} 
\eea
where we have abbreviated $\kappa:=\frac{1\!+\!  v}{l \hs ^3}\ge 0$. \
From $\sigma$ one can immediately obtain $\partial u^z/\partial Z$ via
\be
l\frac{\partial u^z}{\partial Z}=- \frac{1+v}{2\hs^2}\,\sigma.                     \label{Def-dudz}
\ee
(this is dimensionless, as well). To bound $\varepsilon,\sigma$ 
for small $\xi$ we introduce the Liapunov function  
\be
V:=  \varepsilon^2\!+\!\beta\sigma^2,
\ee
where $\beta(Z)\in\RR^+$ is specified below. Clearly  $V(0,Z)=0$.
\ Eq. (\ref{basic1}) implies
\bea
V' &=& \varepsilon\sigma\: 2\left(\beta Kl
\check n \!-\!\kappa\right)  +\sigma\: 2\beta Kl \left(\check n -\! \widetilde{n_0}\right)
 \label{Veq1} 
\eea
and, since $ 2|\varepsilon\sigma|\le V/\sqrt{\beta}$, $ |\sigma|\le \sqrt{V/\beta}$,
we obtain
\bea
V' \le  2A\sqrt{V}+ 2B V \quad\Rightarrow\quad 
\left(\!\sqrt{V}\right)'\le A+B\sqrt{V}, 
\nonumber
\eea
where we have abbreviated  \ $B(\xi,Z):= |\beta Kl\check n \!-\!\kappa|/2\sqrt{\beta }$ \ and introduced some $A(Z)$ such that \ $A\ge Kl\sqrt{\beta }\max\{|\check n\! -\! \widetilde{n_0}|\}$. \
By the comparison principle \cite{Yos66}, $\sqrt{V}\le R$, where $R(\xi,Z)$ is the solution of the   Cauchy problem \ $R'= A\!+\!BR,\:\: R(0,Z)=0$, \ what implies
\bea
|\varepsilon(\xi,Z)|\le\sqrt{V(\xi,Z)}\le R(\xi,Z)=A(Z)\int^\xi_0\!\!\!d\eta\, \exp\left[\int^\xi_\eta\!\!\!d\zeta\, B(\zeta,Z)\right]
%e^{\int\nolimits^\xi_\eta\! d\zeta\, B(\zeta)}
 \label{varepsilon<R} 
\eea
and \ $\sqrt{b}|\sigma|\le R$.  \ Since $R(\xi,Z)$ grows with $\xi$, if \ $R(l,Z)<1$ \ no WBDLPI may involve the $Z$-electrons. 
Choosing \ $\beta =1/M_ul^2$, \  we find for all $\xi\!\in\![0,\tilde\xi_3']$ 
\bea
2l B\,\beta =\left|\frac{1\!+\!  v}{\hs ^3}\!-\! \frac{\check n}{ n_u}\right|\le  \left|\frac{1\!+\!  v}{\hs ^3}\!-\!1\right| \!+\!\left|1\!-\!\frac{\check n}{ n_u} \right|\le |D|\!+\!\delta, \nn [10pt]
\mbox{where}\qquad  D(\xi,Z):=\frac{1\!+\!  v(\xi)}{\left[\hs (\xi,Z)\right]^3}\!-\!1,\quad 
\delta(Z):=1-\frac{n_d(Z)}{n_u(Z)}.       \label{Def-delta}
\eea

In the NR  regime, which is characterized by $v\ll 1$, we have \ $\hs\simeq1$,
 \ $\kappa\simeq1/l$, $\Delta_u/l\ll 1$, \  $D\simeq 0$; \ setting $D=0$ leads by a straightforward computation to
\bea
R(l,Z)\simeq R_{nr}(Z) :=f[r(Z)],\qquad f(r):=
2\left(e^{r/2}-1\right)
\qquad r(Z):=\delta(Z)\,\sqrt{\!M_u(Z)}\,l. \nonumber
\eea
% \bea J''=-K\left(\check n J -\! \widetilde{n_0}\right), \qquad\quad 
%J(0,Z)=0,\quad J'(0,Z)=0.               \label{basic1nr} \eea
$f(r)$ grows with $r\ge 0$ and reaches the value 1 for $r\simeq 0.81$;
therefore the condition %$r(Z)<0.81$, i.e.
\be
r(Z)%=\delta(Z)\,\sqrt{\!M_u(Z)}\,l
=\delta(Z)\,\sqrt{\!M_u(Z)}\,l<0.81 \label{NR-NoWBcond}
\ee
is sufficient to ensure that the $Z$-electrons are not involved in WBDLPI. This  is automatically satisfied if $\sqrt{\!M_u(Z)}\,l<0.81$, because by definition $\delta\le 1$, otherwise it is a very mild  condition on the relative variation $\delta$ of the initial electron density across an interval of length $\hDOu\ll l$; \  
in fact, to violate (\ref{NR-NoWBcond}) one needs a discontinuous, or 
a continuous but very steep, $\widetilde{n_0}(Z)$ with large relative variations
around $Z$, see section \ref{discuss}.

\noindent
Now consider the general  case.
In the interval  $[0,\tilde\xi_3']$ the inequalities $\hsU\ge \hs \ge \hsD\ge 1$ imply
%$|D|\le v$ where $D\ge 0$ and $|D|\le 1$ where $D\le 0$, i.e.
\bea
 \left| D(\xi,Z)\right| & \le & \D(\xi,Z):
= \max\left\{ \frac{1\!+\!  v(\xi)}{\left[\hsD(\xi,\!Z)\right]^3} - 1 \:, \:
1 - \frac{1\!+\!  v(\xi)}{\left[\hsU(\xi,\!Z)\right]^3}\right\}\nn[6pt]
& \le & \tilde v(\xi):=\max\{v(\xi),1\} \\[6pt]
& \le &\max\{\vM,1\}=:\tvM, \nonumber
\eea
where $\vM,\tvM$ are the maxima of $v,\tilde v$. Hence we obtain the bounds
\bea
 \int^{\xi}_\eta\!\!\!d\zeta\, B(\zeta) & \le & \frac{\sqrt{\!M_u}}2\left[(\xi\!-\!\eta) \,\delta
\!+\!\! \int^{\xi}_\eta\!\!\!d\zeta\, \D(\zeta) \right] \nn
& \le &  \frac{\sqrt{\!M_u}}2\left[ (\xi\!-\!\eta)\, \delta
\!+\!\! \int^{\xi}_\eta\!\!\!d\zeta\, \tilde v(\zeta) \right]\nn
& \le &   \frac{\sqrt{\!M_u}}2\left(\tvM\!+\!\delta\right)(\xi\!-\!\eta) ,\nonumber
\eea
which replaced in (\ref{varepsilon<R}) by choosing  \ $A=\sqrt{M_u}\,\delta$ \
respectively imply
\bea
R(\xi,Z)  &   \!\!\le\!\! & \delta(Z)  \sqrt{\!M_u(Z)}\!\!
\int\limits^{\xi}_0\!\! d\eta\, \exp\!\left\{\!\frac{\sqrt{\!M_u(Z)}}2\left[ (\xi\!-\!\eta)\, \delta(Z)
\!+\!\! \int^\xi_\eta\!\!\!\!d\zeta\, \D(\zeta,\!Z) \right]\!\right\} \!=:Q(\xi,\!Z) \qquad \label{NoWBcond2}\\
Q(l,\!Z) &   \!\!\le\!\! & \delta(Z)\sqrt{\!M_u(Z)}  \!\!\int\limits^l_0\!\!\!d\eta\, \exp\!\left\{\!\frac{\sqrt{\!M_u(Z)}}2\left[\!(l\!-\!\eta)\, \delta(Z) \!+\!\! 
\int^l_\eta\!\!\!d\zeta\, \tilde v(\zeta) \!\right]\!\!\right\} \!=:\! Q_1(Z)\quad \label{NoWBcond1} \\
&  \!\!\le\!\! &\frac{2\delta(Z)}{\tvM\!+\!\delta(Z)} \left\{\exp\left[\frac{\tvM\!+\!\delta(Z)}2\sqrt{\!M_u(Z)}\,l\right]-1\right\}=:Q_0(Z)    \label{NoWBcond0}
\eea
(here we have redisplayed the $Z$-argument). $Q_2(Z) \!:=\! Q(l,\!Z) $ is the 
most difficult to compute,   $Q_0(Z)$ is the easiest. We thus arrive at  

\begin{theorem}
Assume that  condition (\ref{ShortPulse1'}) is fulfilled. 
Then no WBDLPI  involves the $Z$-electrons if, in addition, $Q_0(Z)<1$, or at least $Q_1(Z)<1$,
 or at least $Q_2(Z)<1$. If one  of these conditions is fulfilled for all $Z$,
then  WBDLPI  occurs nowhere.
\label{PropNoWB}
\end{theorem}

Consequently, for any fixed pump there is no WBDLPI  if $n_b$ is sufficiently small. 
A simple sufficient condition is given by

\begin{corollary} 
For any pulse (\ref{pump}) there is no WBDLPI if \ \ $K n_b l^2<4\big[\log 2/(1\!+\!\tvM)]^2$ \ and (\ref{ShortPulse1'}) are fulfilled. In particular, it suffices that 
 \ \ $K n_b l^2<\min\left\{4\big[\log 2/(1\!+\!\tvM)]^2,2/(1\!+\!2\Delta_u/l)\right\}$.
\end{corollary}
\bp{}
From $\delta\le 1\le \tvM$ it follows $\delta\,l\,\sqrt{M_u}\le L:=\frac{\tvM\!+\!\delta}2\,l\,\sqrt{M_b}$.
Hence $L<\log 2$ implies $e^L<2$, whence
$$
Q_0=\delta\,l\,\sqrt{M_u}\:\frac{e^L\!-\!1}{L}\le e^L\!-\!1<1,     
$$
which together with (\ref{ShortPulse1'}) implies the first claim. The second follows as (\ref{ShortPulse2})
implies (\ref{ShortPulse1'}).
\ep

\section{Discussion and conclusions}
\label{discuss}

As we have seen, if the inequality (\ref{ShortPulse1}) is fulfilled, then $\tilde\xi_2> l$, 
i.e. the pulse is strictly short (namely, it completely overcomes the $Z$-electrons while their longitudinal displacement is still nonnegative). If at least the inequality (\ref{ShortPulse1'}) is fulfilled, then $\tilde\xi_3> l$,
so that the pulse is  essentially short (namely, the pulse completely overcomes the $Z$-electrons before their longitudinal displacement reaches its first negative minimum), and the inequalities (\ref{sDeltabound1}), (\ref{Hbound}),
(\ref{NoWBcond2}-\ref{NoWBcond0}) apply. 
If in addition one of the conditions of Proposition \ref{PropNoWB} is satisfied, then we can indeed exclude WBDLPI.

As seen, the  more easily computable - but also more difficult to satisfy - condition 
 (\ref{ShortPulse2}) implies (\ref{ShortPulse1}) and also the inequality \
$M_u(Z)\,  l^2\le \frac 2{1\!+\! 2\Delta_u/l} $, which after a substitution in
(\ref{NoWBcond2}-\ref{NoWBcond0}) simplifies the computation of their rhs;
in particular (\ref{NoWBcond0}) becomes
\bea
 Q_0(Z) \le \frac{2\delta(Z)}{\tvM\!+\!\delta(Z)} \left( e^C-1\right)
%\left\{\exp\left[\frac{\tvM\!+\!\delta(Z)} %{\sqrt{2\left[1\!+\!\frac{2\Delta_u}l\right]}}
%{\sqrt{2\left[1\!+\! 2\Delta_u/l\right]}}\right]-1\right\},
=:\tilde Q_0(Z),\qquad C:=\frac{\tvM\!+\!\delta(Z)}
%{\sqrt{2\left[1\!+\!\frac{2\Delta_u}l\right]}}
{\sqrt{2\left[1\!+\! 2\Delta_u/l\right]}}.   \label{NoWBcond0'}
\eea
Therefore  (\ref{ShortPulse2}) and $\tilde Q_0(Z)<1$ provide a sufficient condition to exclude WBDLPI, as well.

Note that, in the above conditions, several dimensionless numbers  characterizing the input data, viz.
\  $\tvM,\, \Delta_u/l,\, G_b^2=M_bl^2,\, M_ul^2,\, M_dl^2,\, \delta$, \  and possibly also $\hs_u,\, M_u'l^2$,  play a key role 
in the main inequalities of the present paper;
%of  propositions \ref{PropBounds-sDeltaH}, \ref{PropNoWB}, as well as in  (\ref{ShortPulse1}),   (\ref{ShortPulse2}), (\ref{hbound}) 
therefore their computation represents the first step to check whether/where such conditions
are fulfilled or violated.

In the NR regime %, by a little algebra one easily shows that 
(\ref{NR-NoWBcond}) is equivalent to either inequality\footnote{Ineq. (\ref{NR-NoWBcond}) amounts to
$n_u\!-\!n_d<\sqrt{2p\,n_u}$, i.e.  (\ref{NR-NoWBcond1}a); taking the square one obtains the equivalent inequality
$n_u^2\!-\!2n_u(n_d\!+\!p)\!+\!n_d^2\!<\!0$, which is fulfilled if
$n_-\!<\!n_u\!<\!n_+$, where $n_\pm\!:=\!n_d\!+\!p\!\pm\!\sqrt{(n_d\!+\!p)^2\!-\!n_d^2}
=n_d\!+\!p\!\pm\!\sqrt{p^2\!+\!2pn_d}$ solve the equation
$x^2\!-\!2x(n_d\!+\!p)\!+\!n_d^2=0$ in the unknown $x$; the left inequality is automatically satisfied because
$n_u\ge n_d>n_d^2/n_+=n_-$. Dividing the inequality $n_u<n_+$
by $p$ we obtain (\ref{NR-NoWBcond1}).}
\bea
\frac{n_d}p>\frac{n_u}p-\sqrt{\frac{2n_u}p}\quad\Leftrightarrow\quad
\frac{n_u}p<1+\frac{n_d}p+\sqrt{1+2\frac{n_d}p}, \qquad p:=\frac{(0.81)^2}{2Kl^2}.
\label{NR-NoWBcond1}
\eea
If $\widetilde{n_0}$ grows in $[0,\bar Z\!+\!\Delta_u]$, also $q(z):=\widetilde{n_0}(z)/p$ does, and for all 
$\in[0,\bar Z]$ the previous conditions become
\bea
q(Z)>q\big(Z\!+\!\Delta_u\big)-\sqrt{2q\big(Z\!+\!\Delta_u\big)}\quad\Leftrightarrow\quad
q\big(Z\!+\!\Delta_u\big)<1\!+\!q(Z)\!+\!\sqrt{1\!+\!2q(Z)}.
\label{NR-NoWBcond2}
\eea
This is fulfilled e.g. if  in $]0,\bar Z\!+\!\Delta_u]$ $\widetilde{n_0}$ is continuous (without
excluding $\widetilde{n_0}(0^+)>0$),  at least piecewise $C^1$, 
and $0\le \frac{dq(z)}{dz}\: \Delta_u < 1+\!\sqrt{1\!+\!2q(Z)}$ \ for all $Z\in[0,\bar Z]$
and $z\in[Z,Z\!+\!\Delta_u]$.

Now we impose that  $\widetilde{n_0}$ is continuous in $]-\infty,\bar Z\!+\!\Delta_u]$ and reaches a given value $\bar n>0$ at $Z=\bar Z$ while respecting (\ref{NR-NoWBcond2}).
We compare the minimum $\bar Z$ for a linear and a quadratic $\widetilde{n_0}$; note that $dq/dZ$ for
the former violates the above bound at $Z=0$. We 
find\footnote{In fact, if $\widetilde{n_0}=n_1$ then (\ref{NR-NoWBcond2}b) becomes 
$\bar n \Delta_u/p{\bar Z}<1\!+\!\sqrt{1\!+\!2\bar n Z/p{\bar Z}}$ for all $Z$; the rhs is lowest  for
$Z=0$, whereby the inequality becomes (\ref{Z_1bound}b), as claimed. \ \ 
If $\widetilde{n_0}=n_2$ then (\ref{NR-NoWBcond2}a) becomes  the condition
\bea
F(Z):=\sqrt{\frac{2p}{\bar n}}\bar Z(Z\!+\!\Delta_u)\!-\!\Delta_u^2\!-\!2\Delta_uZ>0; 
\nonumber
\eea
this is of  first degree in $Z$, hence is fulfilled for all $Z\in[0,\bar Z]$ if it is for $Z=0,\bar Z$; the quadratic polynomial $F(\bar Z)$ in $\bar Z$ 
is positive if $\bar Z>Z_2$, as claimed, because $Z_2$ is the positive solution of the equation $F(z)=0$ in the unknown $z$, and $\bar Z>Z_2$ automatically makes also $F(0)>0$.
}
\bea
\widetilde{n_0}(Z)=n_1(Z):=\theta(Z)\bar n \frac Z{\bar Z}\:\: &\mbox{fulfills (\ref{NR-NoWBcond2}) if}& \:\:
\frac{\bar Z}{\Delta_u}> \frac{\bar n}{2p}=\frac{K\bar n l^2}{(0.81)^2}=:\frac{\bar Z_1}{\Delta_u} \label{Z_1bound}\\
\widetilde{n_0}(Z)=n_2(Z):=\theta(Z)\bar n \frac {Z^2}{\bar Z^2}\:\: &\mbox{fulfills (\ref{NR-NoWBcond2}) if}&
\:\:\frac{\bar Z}{\Delta_u}>\sqrt{\frac{\bar n}{2p}}\!+\!\sqrt{\frac{\bar n}{2p}\!+\!\frac 14}\!-\!\frac 12%=\Delta_u\frac{\sqrt{K\bar n}l}{0.405} 
=:\frac{\bar Z_2}{\Delta_u}
\eea
 (here $\theta$ is the Heavisde step function).
If $\sqrt{\!K\bar n}\,l$ is considerably larger than 1, then
 $\bar Z_1$  is considerably larger than  $\bar Z_2$;   in particular, assuming (\ref{MaxTransfer}) 
with $n_b=\bar n$, i.e. $G_b\sim \pi$, yields $Z_1\sim 15.04\Delta_u$,  $Z_2\sim 6.78\Delta_u$. Therefore choosing $\bar Z\in]Z_2,Z_1[$ we can exclude  WBDLPI   adopting $\widetilde{n_0}(Z)=n_2(Z)$, but not $\widetilde{n_0}(Z)=n_1(Z)$.
Such a result is relevant for LWFA experiments, which usually fulfill (\ref{MaxTransfer}).
From the physical viewpoint, it allows one to exclude WBDLPI because: i)
the density  $\widetilde{n_0}(Z)$ obtained just outside the nozzle  of a
supersonic gas jet (orthogonal to the $\vec{z}$) tipically is $C^1(\RR)$ with 
$\widetilde{n_0}(0)=0=\frac{d\widetilde{n_0}}{dZ}(0)$ (see e.g. fig. 2 in \cite{HosEtAl02}, or fig. 5 in  \cite{VeiEtAl11}), 
and therefore is closer to  type
$n_2(Z)$ than to type $n_1(Z)$; ii) by causality, the effects of a pulse
with a finite spot radius $R$
near its symmetry axis $\vec{z}$ are the same as with a plane wave ($R=\infty$), at least for small $\xi$.
%This result hints at a more general lesson, namely
From the viewpoint of mathematical modeling, it suggests that it makes a big difference to describe the edge
of the plasma by $n_1$ or by $n_2$: in the first case we can
correctly  predict the plasma evolution only by kinetic theory and PIC codes, while in
the second we can do also by a hydrodynamic description and  (less computationally demanding)
multifluid %simulation 
codes.

If we allow a discontinuous (in $Z=0$) linear Ansatz  $\widetilde{n_0}(Z)=\theta(Z)\bar n[a+(1\!-\!a)Z/\bar Z]$ 
($0\!< \!a\!\le 1$), then (\ref{NR-NoWBcond2}) is fulfilled if
$\bar Z> \bar n (1\!-\!a)\Delta_u/p[1\!+\!\sqrt{1\!+\!2a\bar n/p}]$, which is again smaller than $Z_1$.

\medskip
Although our results apply to all $\Bep$ with support contained in $[0,l]$, regardless of their Fourier analysis, in most applications one deals with a modulated monochromatic wave\footnote{
The elliptic polarization in (\ref{modulate}) is ruled by $\psi,\varphi_1,\varphi_2$; it  reduces to a linear one in the direction 
of ${\bm a}:=\bi\,\cos\psi+\bj\,\sin\psi$ if
$\varphi_1=\varphi_2$, to a circular one if $|\cos\psi|=|\sin\psi|=1/\sqrt{2}$ and $\varphi_1=\varphi_2\pm\pi/2$.},
\be
\Bep\!(\xi)\!=\!\underbrace{\epsilon(\xi)}_{\mbox{modulation}}
\underbrace{[\bi \cos\psi\,\sin (k\xi\!+\!\varphi_1)\!+\!\bj \sin\psi\sin (k\xi\!+\!\varphi_2)]}_{\mbox{carrier wave $\Be_o^{{\scriptscriptstyle \perp}}\!(\xi)$}},
 \label{modulate}
\ee
where $\bi=\nabla x$,  $\bj=\nabla y$. If, as it follows from (\ref{Lncond}),
$Kn_b \lambda^2\ll1$ ($\lambda=2\pi/k$ is the wavelength), i.e. the plasma is underdense, then the relative 
variations  of $\hD (\xi,Z)$ (and $\hze(\xi,Z)$) in a $\xi$-interval of length $\le\lambda$ are much smaller than those of $\hbx_e^{\scriptscriptstyle \perp}(\xi)$, and those
$\hs (\xi)$   even smaller; in fact, as $\hs \!>\!0$, $v\!\ge\! 0$, the 
integral in (\ref{sDelta}a) averages the fast variations of $v$ to yield much smaller relative variations of $\hD $,  and the first 
integral in (\ref{sDelta}b) averages the residual small 
 variations of $\widetilde{N}[\hze(\xi)]$ to yield an essentially smooth $\hs (\xi)$, see e.g. fig. \ref{graphs}.
On the contrary, $\Bap(\xi),\hbx_e^{\scriptscriptstyle \perp}(\xi)$ vary fast as $\Bep(\xi)$.
Under rather general assumptions  (see the appendix) \cite{Fio18JPA}
\be
\Bap(\xi)= -   \frac {\epsilon(\xi)}k \:\Bep_p\!(\xi)+O\left(\frac 1 {k^2}\right)
 \simeq -   \frac {\epsilon(\xi)}k \,\Bep_p\!(\xi),              \label{slowmodappr}
%,\qquad\quad \Bep_p\!(\xi) := -  \frac {\Bep_o{}'\!(\xi)}k=\Bep_o(\xi\!+\!\pi/2k)
\ee
where $\Bep_p\!(\xi)\! :=\!\Bep_o(\xi+\lambda/4)$, and similarly for other integrals with modulated integrands; in the appendix we  recall upper bounds  for the remainders $O(1/k^2)$. If  $|\epsilon'|\!\ll\! |k\epsilon|$  (slow modulations) 
% - like the ones characterizing  conventional applications (radio broadcasting, ordinary laser beams, but also most )  -
%($\lambda |\epsilon'|%\!\le\!\delta
%\!\ll\! |\epsilon|$, i.e. the modulating amplitude $\epsilon$ does not vary significantly over the wavelength 
%$\lambda\!:=\!2\pi /|k|$,  almost everywhere),
the right estimate is  very good, and $v$ can be approximated very well by
$v\simeq \Bep_p{}^2  \big(e\epsilon /kmc^2\big)^2$.
%Consequently, if $\epsilon(\xi)$ goes to zero also as $\xi\!\to\!\infty$, then 
%$\Ba\!^{{\scriptscriptstyle\perp}}(\xi), v(\xi)$ approximately do  as well.
%See the graphs of the examples below.
For the reasons mentioned above, replacing $v$ by  its 
(approximated) average over a cycle,
\be
v_a(\xi):= \frac 12  \left(\frac{e\,\epsilon(\xi)}{kmc^2}\right)^2, 
\label{cycle-average}
\ee 
has only a small effect on  $\hD $ and almost no effect on  $\hs $, $V$, 
%solving (\ref{heq1}) %[or, equivalently, (\ref{sDelta})] and (\ref{Veq1}), 
and similarly on the functions $\hDO,\hsU,...$ introduced 
in  sections \ref{DsHbounds}, \ref{Jbound} to bound    $\hD ,\hs ,V$; but it simplifies their %numerical 
computation a lot.  As a consequence, the bounds  (\ref{sDeltabound1}),  (\ref{Hbound}) and 
 (\ref{hbound}), as well as the short pulse conditions (\ref{ShortPulse1}),  (\ref{ShortPulse1'}) and the
no-WB conditions  of proposition \ref{PropNoWB},  
remain essentially valid also if in computing the bounds we replace $v$ by $v_a$.

\begin{figure}
\includegraphics[width=16cm]{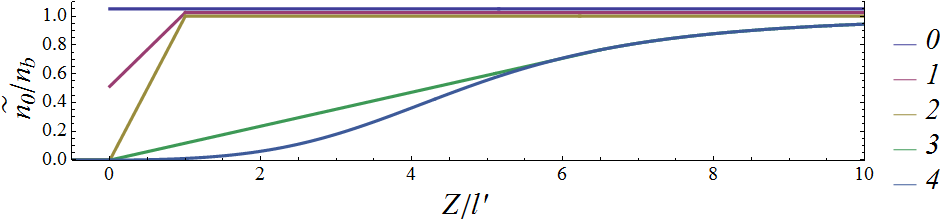}
\caption{Plots of the ratios $\widetilde{n_0}/n_b$ for the following initial densities: \\ 0)  $\widetilde{n_0}(z)=n_b\,\theta(z)$. 
\\ 1) $\widetilde{n_0}(z)=\frac {1}2n_b\,\theta(z)\left[1\!+\!\theta(l'\!-\! z)\,z/l' \!+ \theta(z\!-\! l')\right]$.
\\  2) $\widetilde{n_0}(z)=n_b\,\left[\theta(z)\, \theta(l'\!-\! z)\,z/l' \!+ \theta(z\!-\! l')\right]$.
\\  3) $\widetilde{n_0}(z)=n_b\,\left\{\frac z{\bar z}\,\frac {f(\bar z)}{1\!+\!f(\bar z)}
\theta(z)\theta(\bar z\!-\!z)%\,f(\bar z)/[1\!+\!f(\bar z)]\,z/\bar z
+\theta(z\!-\!\bar z)\,\frac {f(z)}{1\!+\!f(z)}%\,f(z)/[1\!+\!f(z)]
\right\}$, where   $f(z):=(0.1\,z/l')^2+(0.2\,z/l')^4$ and $\bar z=6.5l'$; 
this grows as $z$ for $z\le \bar z$ and coincides with the next one for $z>\bar z$.
\\ 4) $\widetilde{n_0}(z)=n_b\,\theta(z)\frac {f(z)}{1\!+\!f(z)}$.  %the concave completion 
}
\label{InDensityPlots}
\end{figure}

\medskip
We illustrate the  results obtained so far considering a  pulse (\ref{modulate}) with a linear polarization 
(e.g. $\psi=0$) and a modulation of Gaussian type, except that it is cut-off outside the support $0\le\xi\le l$:
\be
\frac{e}{kmc^2}\,\epsilon(\xi)=a_0\:\exp\left[-\frac{(\xi\!-\!l/2)^2}{l'{}^2}2\log 2\right] \, \theta(\xi) \:\theta(l-\xi);
\ee
$l'$  is the  {\it full width at half maximum}  of the intensity $I$ of the electromagnetic (EM) field. More precisely, we adopt the pulse plotted in
fig. \ref{graphs}a, which   has the maximum at $\xi=l/2$ and $a_0=1.3$; this  yields a moderately relativistic electron dynamics and
$\Delta_u\equiv\hDO(l)\simeq 0.45 l'$. In fig. \ref{graphs}b we have plot the associated $v$ and $v_a$.
We perform all computations and plots running our specifically 
designed programs  using an “off the shelf,” general-purpose numerical package on a common notebook for a time lapse bewteen several seconds and several minutes. 
  Let us compare the impact of such a pulse
 on the  density profiles plotted in  fig. \ref{InDensityPlots}. %in the interval $0\le z\le 12 l'$:

\begin{figure}[htbp]
%.~\hskip-.3cm
\begin{minipage}{.49\textwidth}
\includegraphics[width=7.8cm]{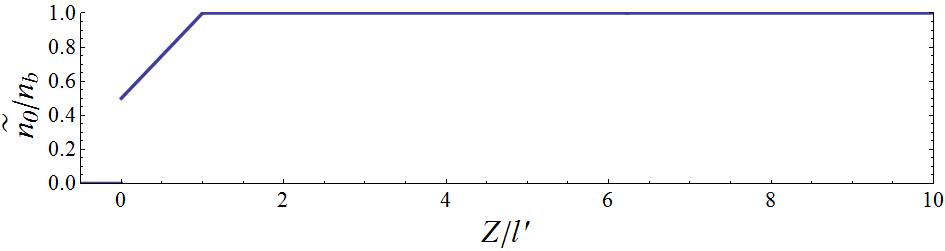}\\
\includegraphics[width=8.0cm]{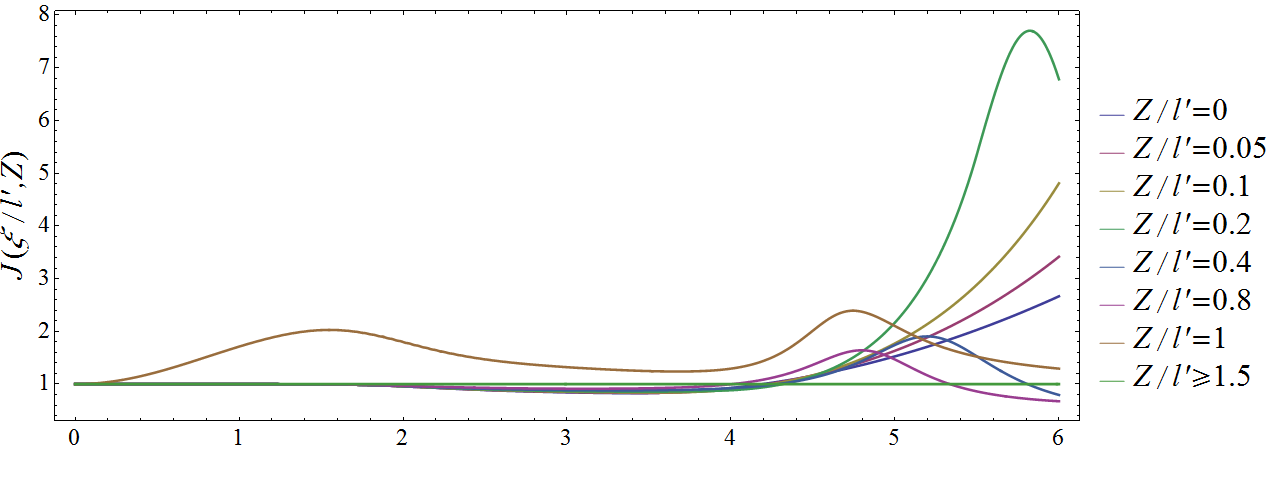}\\
\includegraphics[width=8.0cm]{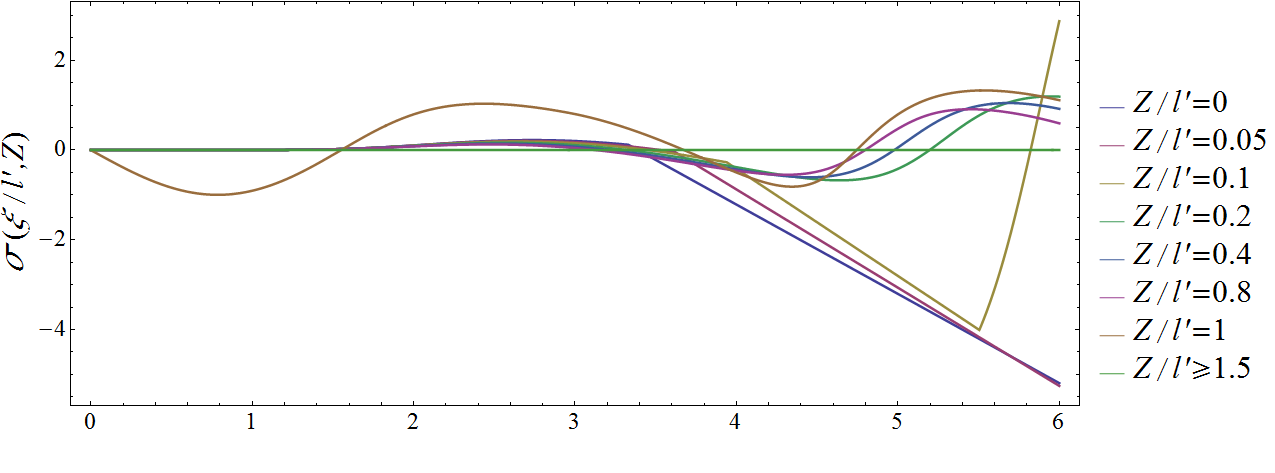}\\
\includegraphics[width=8.0cm]{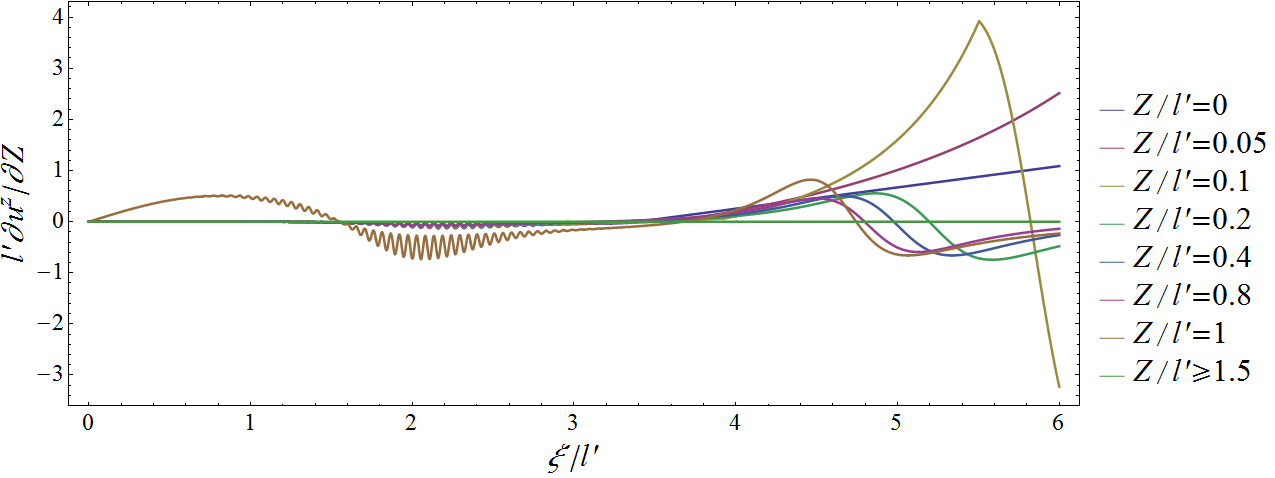}\\
\includegraphics[width=8.0cm]{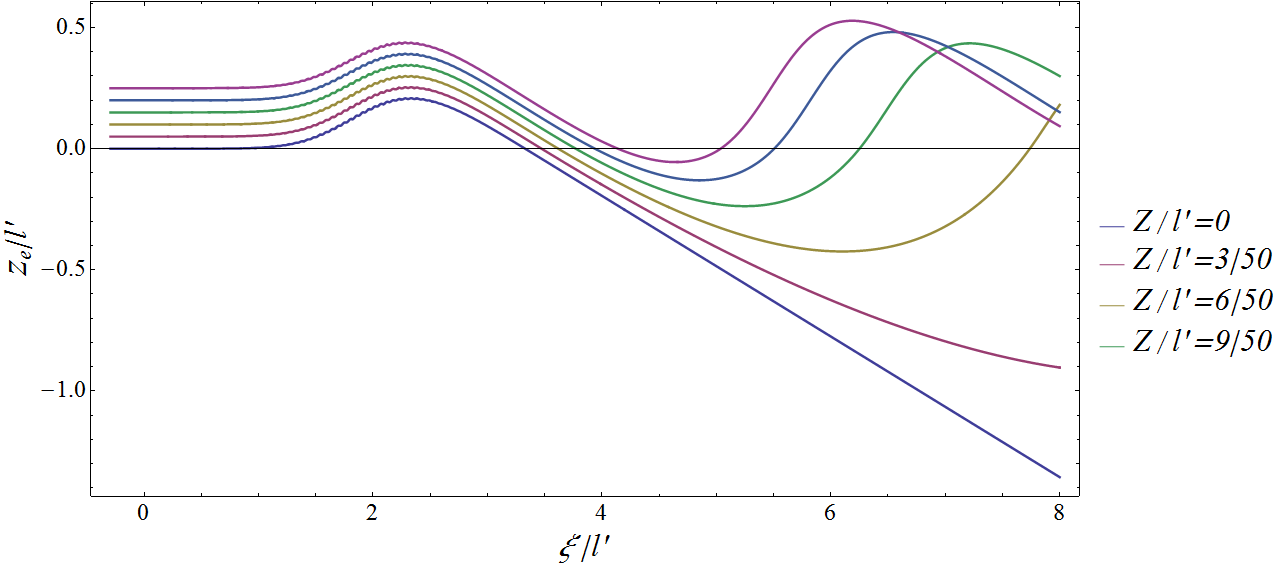}
\end{minipage}
\hfill
\begin{minipage}{.49\textwidth}
\includegraphics[width=7.8cm]{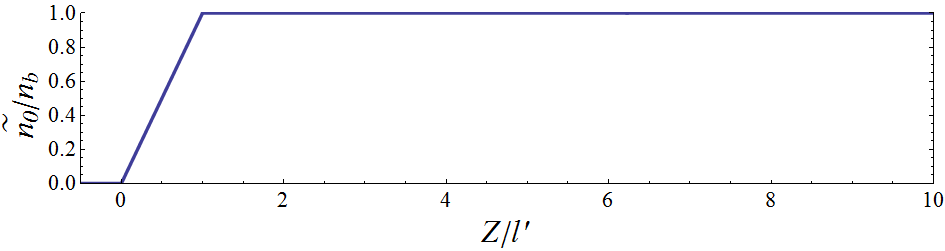}\\
\includegraphics[width=8.0cm]{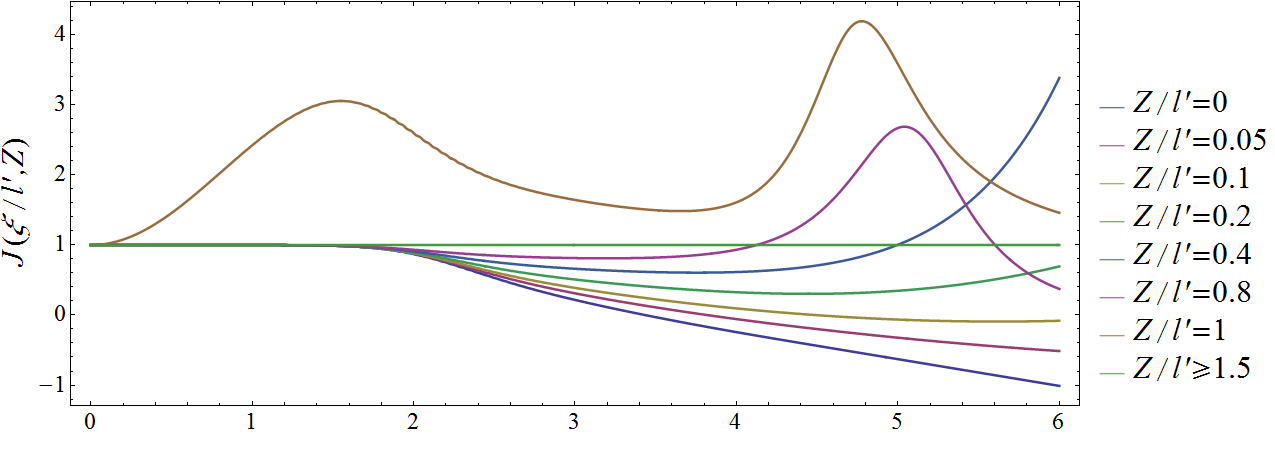}\\
\includegraphics[width=8.0cm]{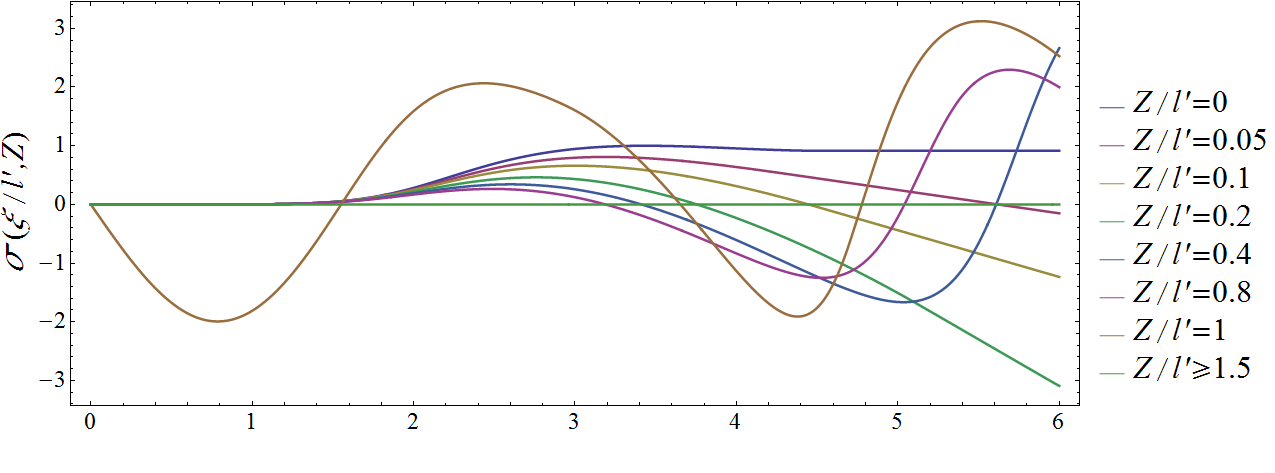}\\
\includegraphics[width=8.0cm]{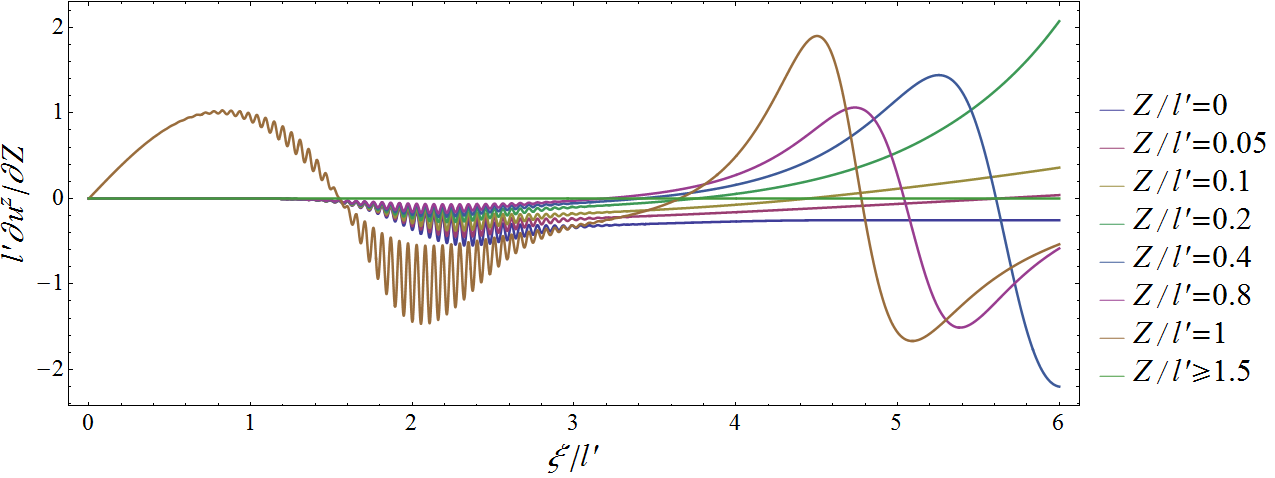}\\
\includegraphics[width=8.0cm]{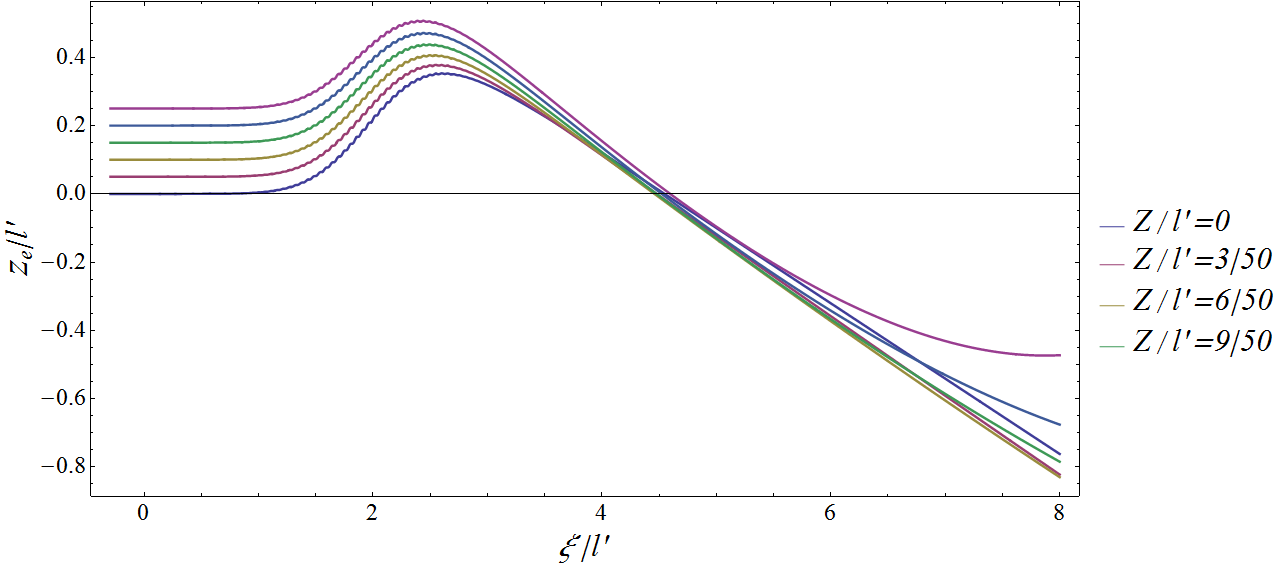}
\end{minipage}
\caption{%Up to down:  
The initial electron densities 1), 2) of fig. \ref{InDensityPlots} (first line; respectively left, right) and, below, the
corresponding plots of $J,\sigma,l'\partial u^z/\partial z$ vs. $\xi$ during the interaction with the  pulse of fig. \ref{graphs}, for a few sample values of $Z$.
As we see, the right $J$ keeps positive for $\xi<l$ and all $Z$, 
while the left $J$ becomes negative for  very small $Z$
and $\xi\lesssim l$; correspondingly, the right worldlines do not intersect, while the right ones do (see the down $z_e$-graphs).}
\label{J-sigma-dudz_Homlin-LinInhom-const-1}
\end{figure}

\begin{figure}[htbp]
%.~\hskip-.3cm
\begin{minipage}{.49\textwidth}
\includegraphics[width=7.3cm]{InitialDensityPlot_LinInhom-const-1}\\
\includegraphics[width=8cm]{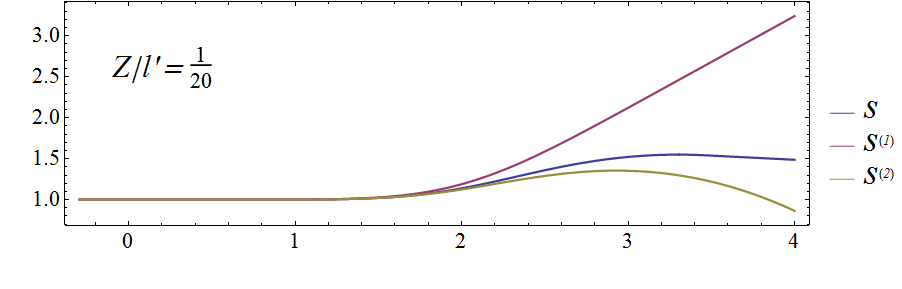}\\
\includegraphics[width=8cm]{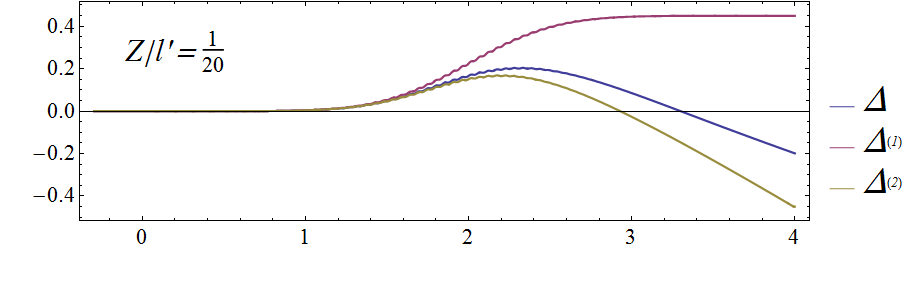}\\
\includegraphics[width=7.9cm]{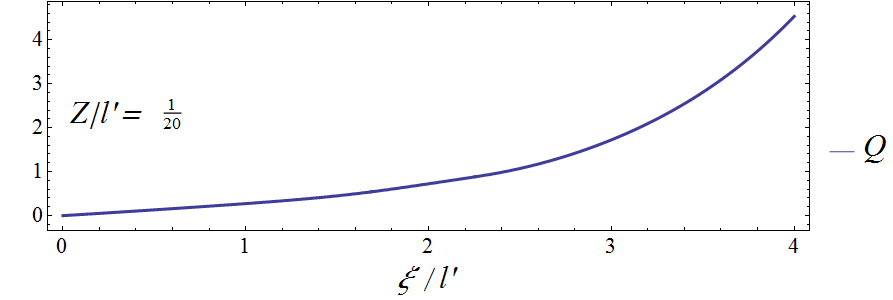}\\
\includegraphics[width=8cm]{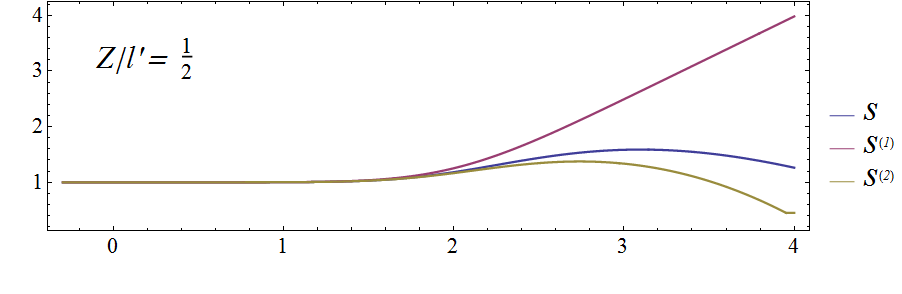}\\
\includegraphics[width=8cm]{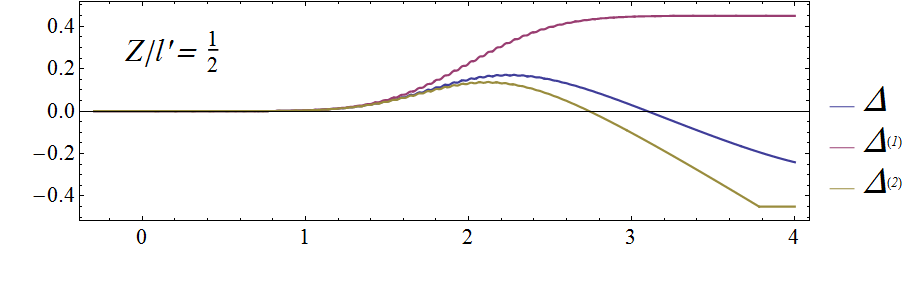}\\
\includegraphics[width=7.9cm]{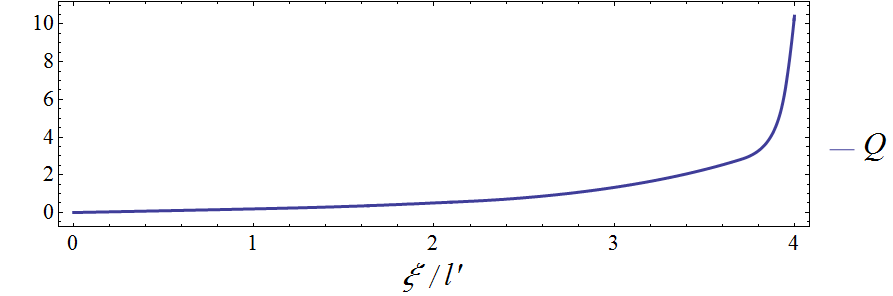}
\end{minipage}
\hfill
\begin{minipage}{.49\textwidth}
\includegraphics[width=7.3cm]{InitialDensityPlot_HomLin-const-1}\\
\includegraphics[width=8cm]{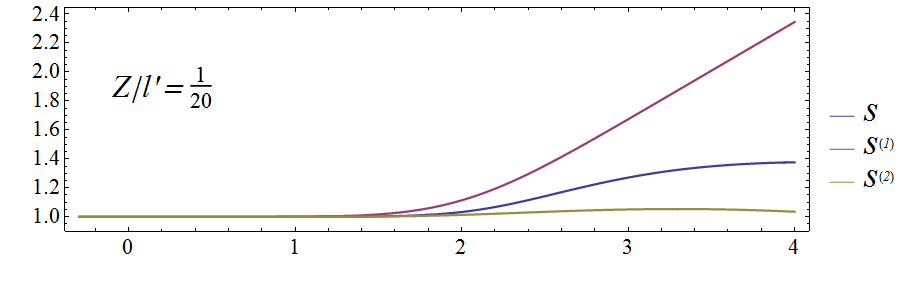}\\
\includegraphics[width=8cm]{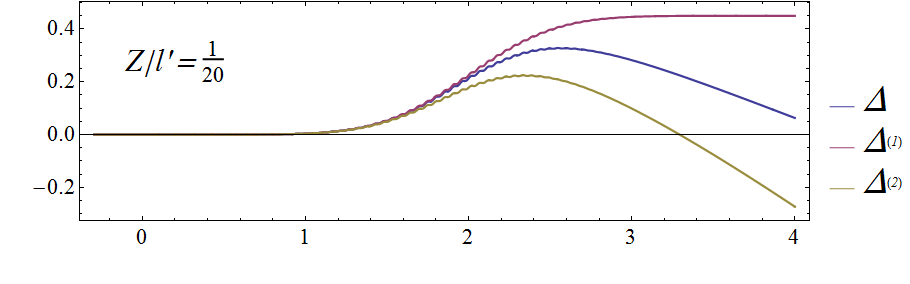}\\
\includegraphics[width=7.9cm]{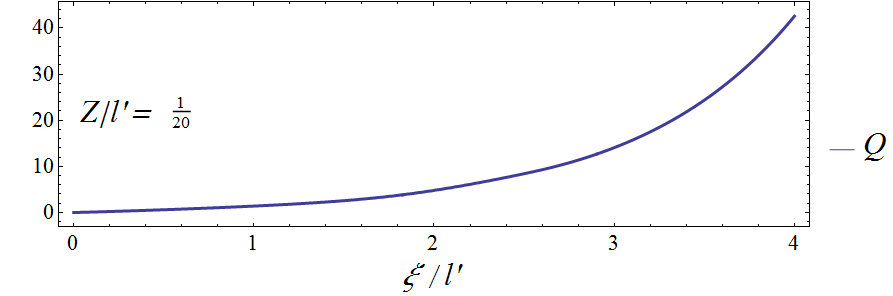}\\
\includegraphics[width=8cm]{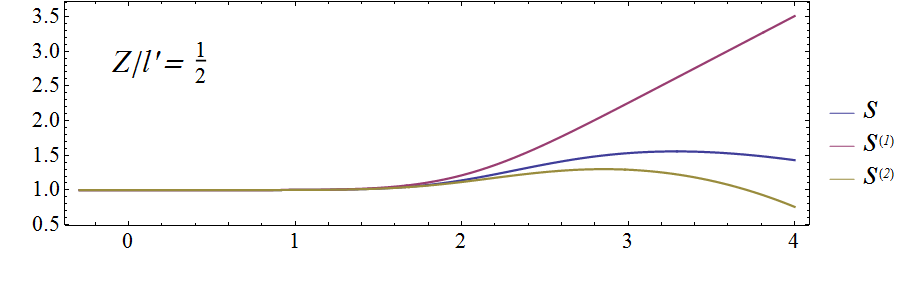}\\
\includegraphics[width=8cm]{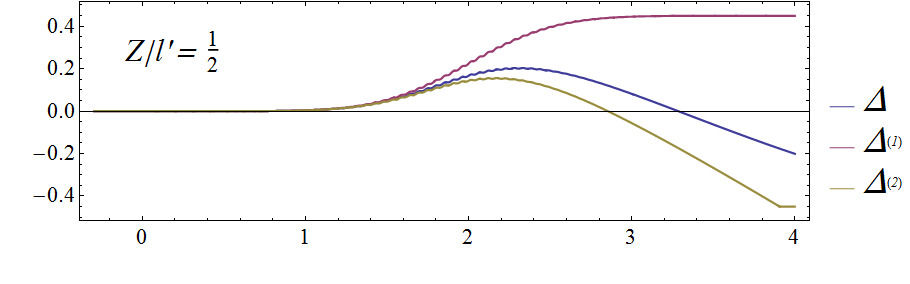}\\
\includegraphics[width=7.9cm]{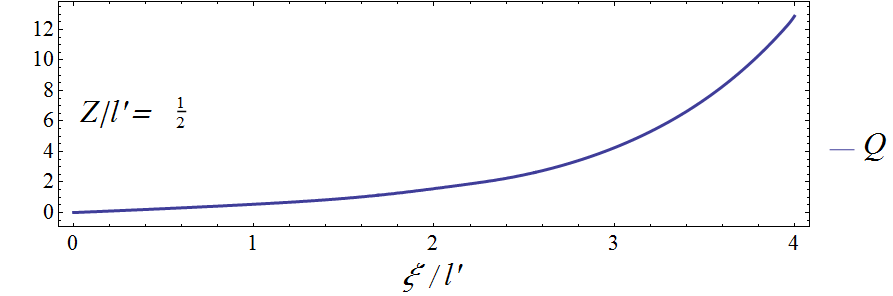}
\end{minipage}%
\caption{%Up to down:  
The initial electron densities 1), 2) of fig. \ref{InDensityPlots} (first line; respectively left, right);
below, assuming $n_b=2\times 10^{18}$cm$^{-3}$, 
we plot the corresponding  $\hs,\hD$,  their upper and lower bounds
$\hsU,\hsD,\hDU,\hDD$ and the function $Q$, vs. $\xi$  during the interaction with the  pulse of fig. \ref{graphs},
  for the same sample values  $Z=l'/20$   and  $Z=l'/2$ of $Z$. The values $Q_2(Z) \!:=\! Q(l,\!Z)$ can be read off the plots. 
As we can see, the bounds are much better for the density 1); the values $Q_2(Z) \lesssim 1$ are consistent
with all worldlines intersecting rather far from the laser-plasma interaction spacetime region. Whereas the large value of $Q_2(Z)$ for  the density 2)
is an indication that some worldlines intersect within, or not far from, the laser-plasma interaction spacetime region. 
Our computations lead also to $Q_0(l'/20)=10.64$, $Q_0(l'/2)=$ with the density 1),
 $Q_0(l'/20)=88.53$, $Q_0(l'/2)=32.35$  with the density 2).}
\label{graphs-bounds_Homlin-LinInhom-const-1}
\end{figure}

\begin{figure}[htbp]
%.~\hskip-.3cm
\begin{minipage}{.49\textwidth}
\includegraphics[width=7.8cm]{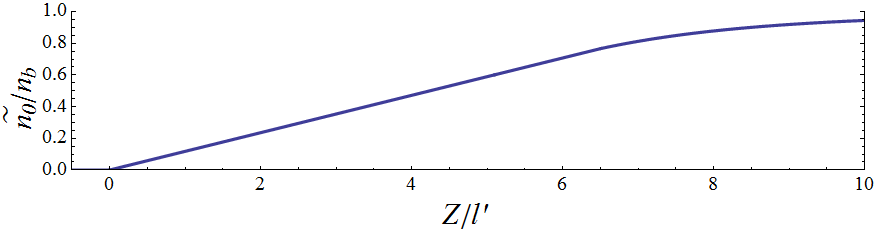}\\
\includegraphics[width=8.0cm]{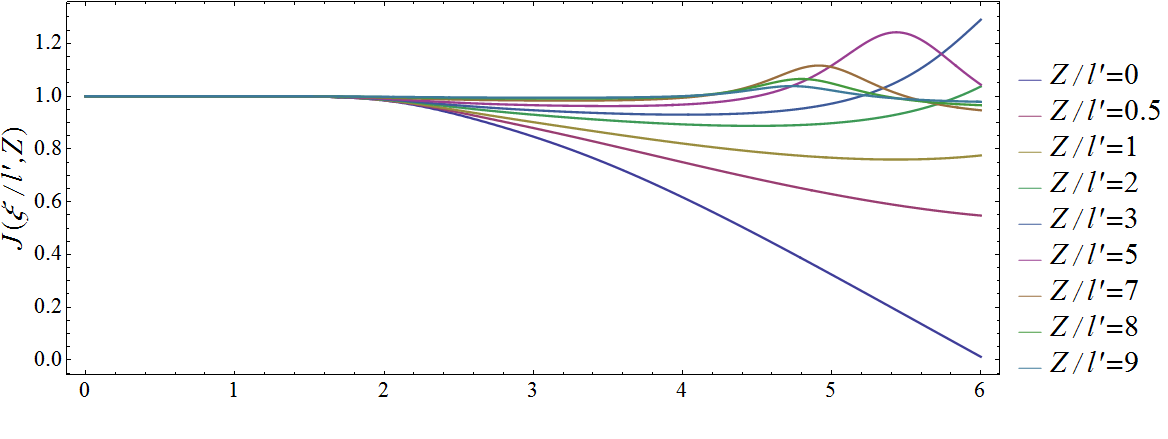}\\
\includegraphics[width=8.0cm]{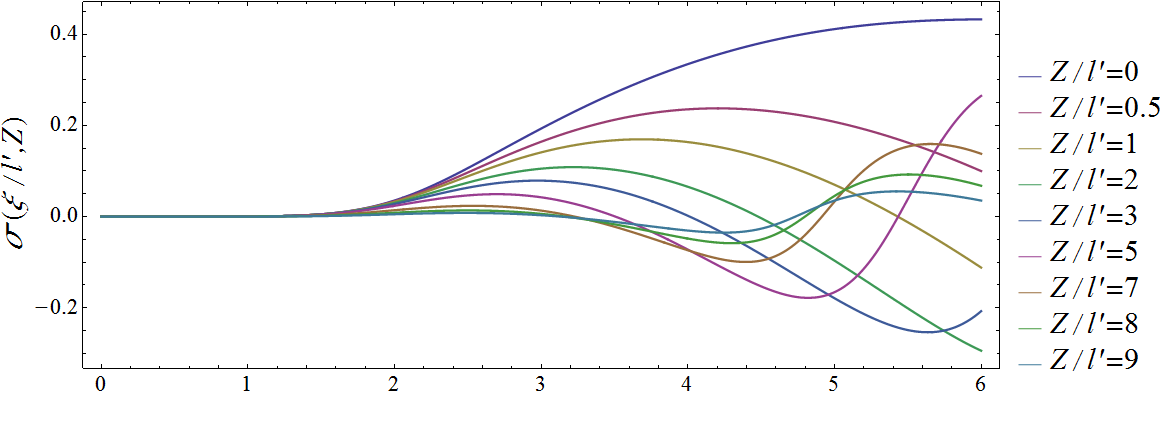}\\
\includegraphics[width=8.0cm]{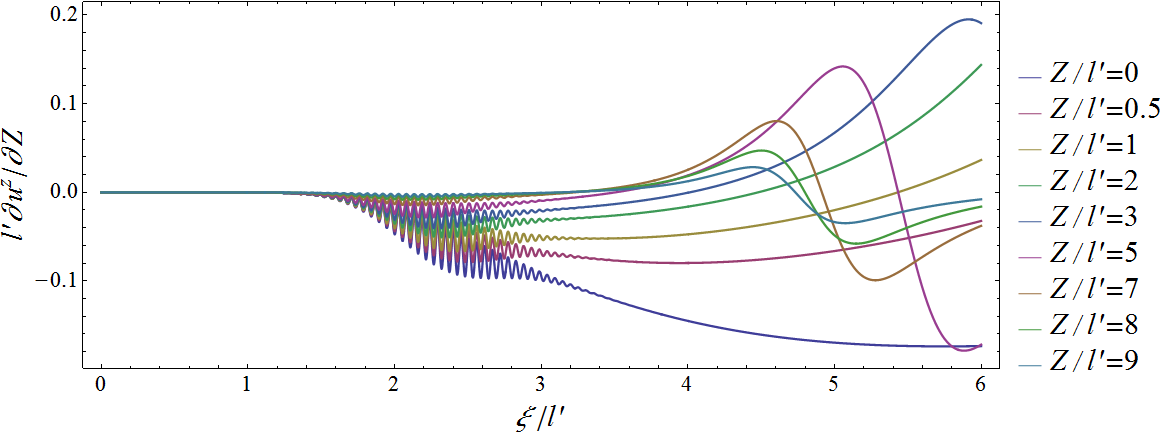}\\
\includegraphics[width=8.0cm]{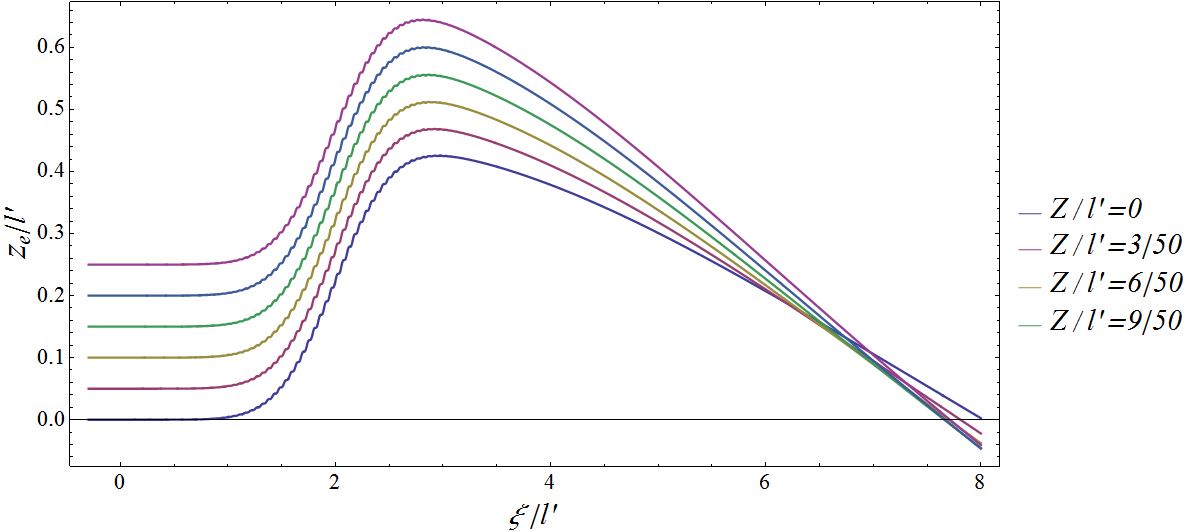}
\end{minipage}%
\hfill
\begin{minipage}{.49\textwidth}
\includegraphics[width=7.8cm]{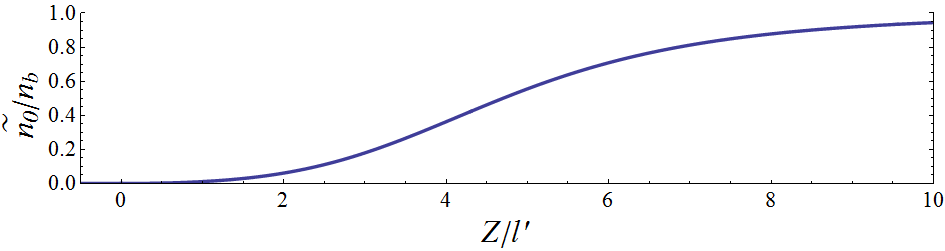}\\
\includegraphics[width=8.0cm]{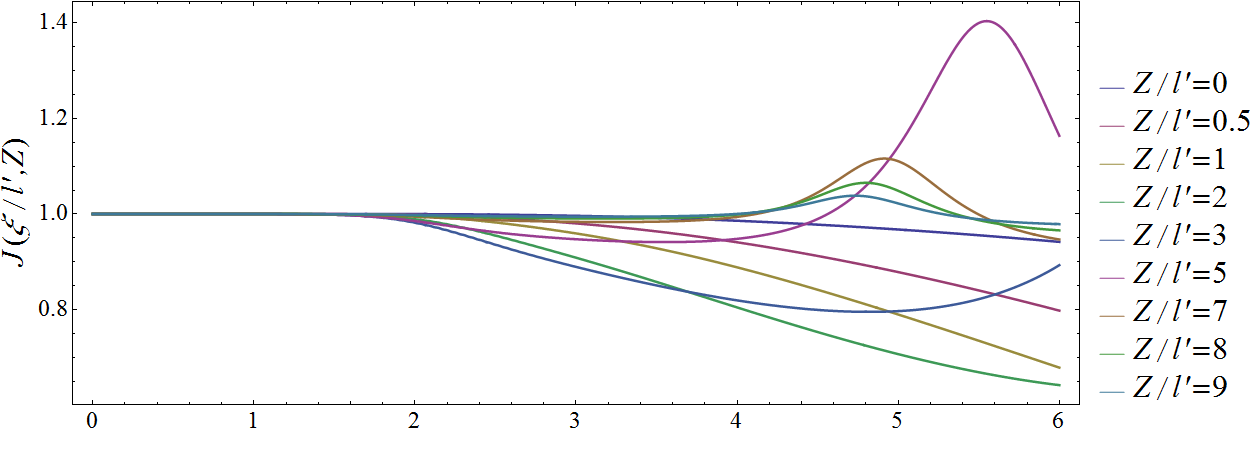}\\
\includegraphics[width=8.0cm]{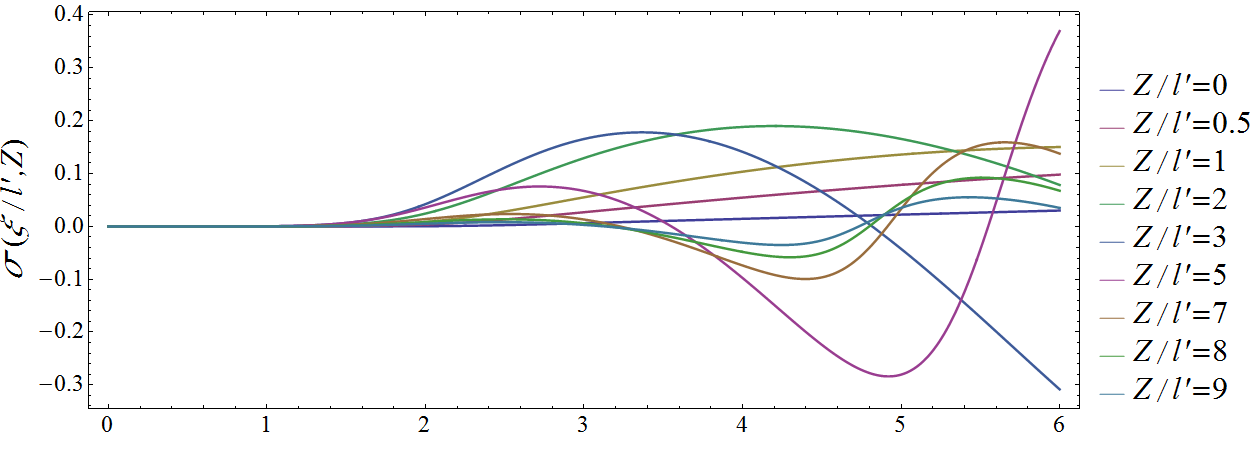}\\
\includegraphics[width=8.0cm]{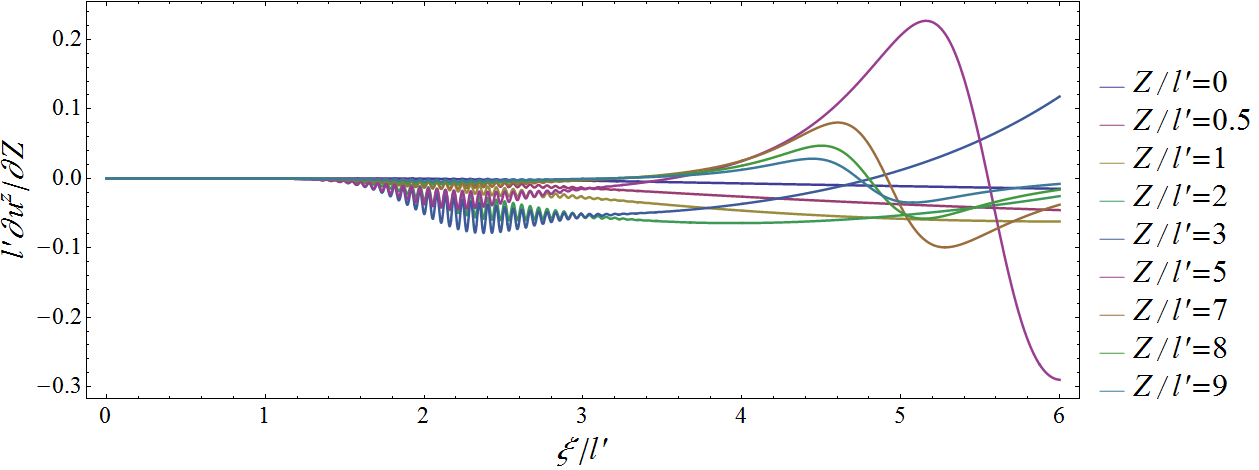}\\
\includegraphics[width=8.0cm]{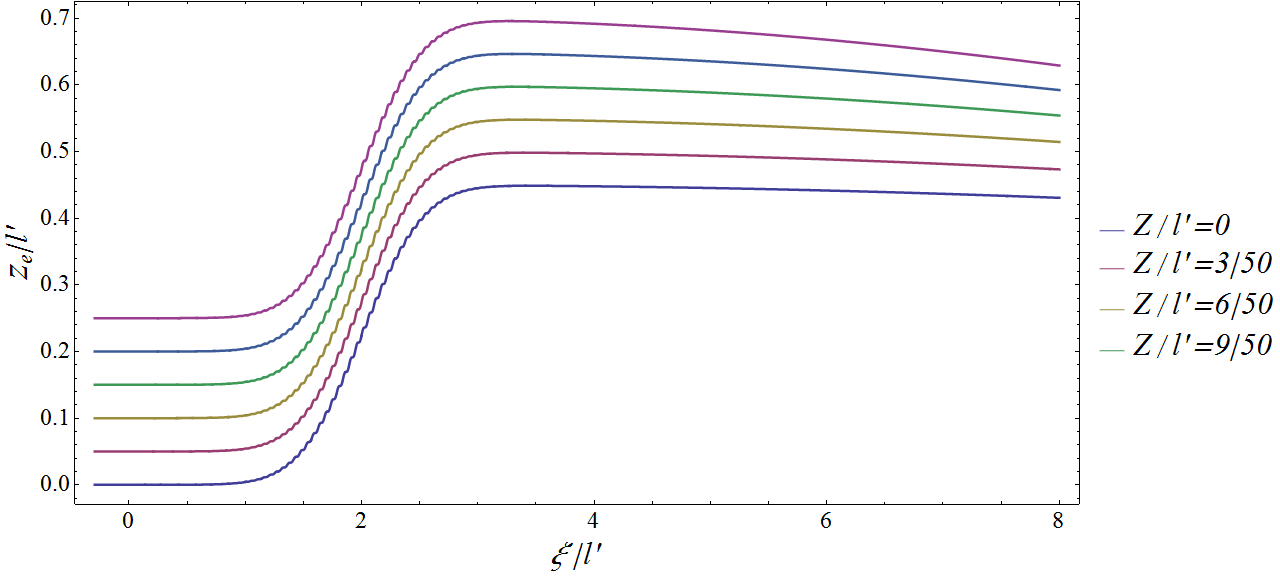}
\end{minipage} 
\caption{%Up to down:  
The initial electron densities 3), 4) of fig. \ref{InDensityPlots} (respectively left, right) with 
$n_b=2\times 10^{18}$cm$^{-3}$, and below the corresponding plots of 
$J,\sigma,l'\partial u^z/\partial z$ during the interaction with the  pulse of fig. \ref{graphs},
for a few sample values of $Z$.
 As we can see,  $J$ keeps positive at least for all $\xi\in[0,2l]$ if
the density is of type 4) (which grows as $Z^2$ for $Z\sim 0$), whereas it becomes negative for $\xi\sim 6.5 l'$
and small $Z$ 
 if the density is of type 3),
 (which grows as $Z$ for $Z\sim 0$); correspondingly, the right worldlines do not intersect, while the left ones do (see the down $z_e$-graphs).
}
\label{J-sigma-dudz_Homlin-Nonlin}
\end{figure}

\begin{figure}[htbp]
%.~\hskip-.3cm
\begin{minipage}{.49\textwidth}
\includegraphics[width=7.3cm]{InitialDensityPlot_HomLin-Nonlin_nb2per10alla18}\\
\includegraphics[width=8cm]{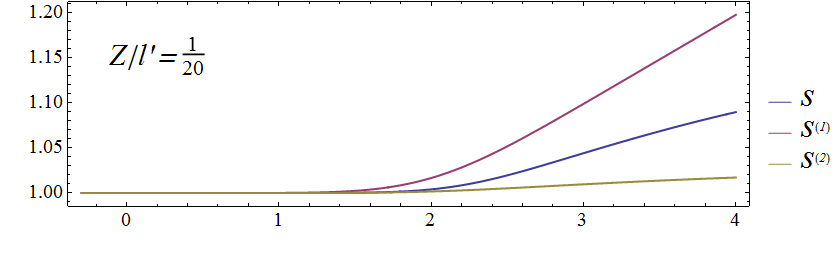}\\
\includegraphics[width=8cm]{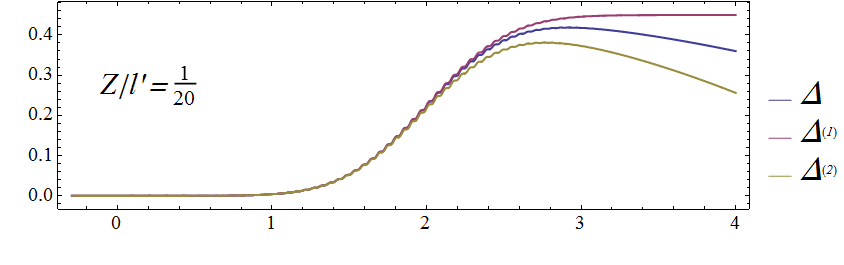}\\
\includegraphics[width=7.9cm]{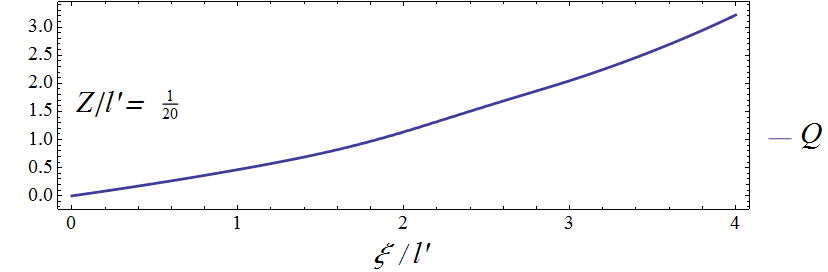}\\
\includegraphics[width=8cm]{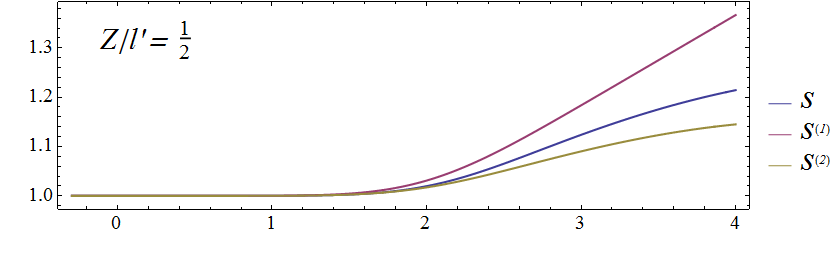}\\
\includegraphics[width=8cm]{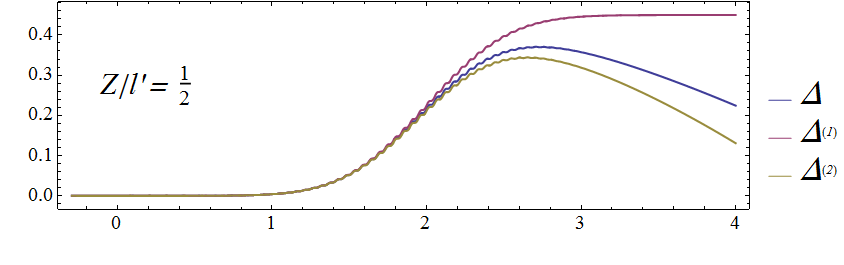}\\
\includegraphics[width=7.9cm]{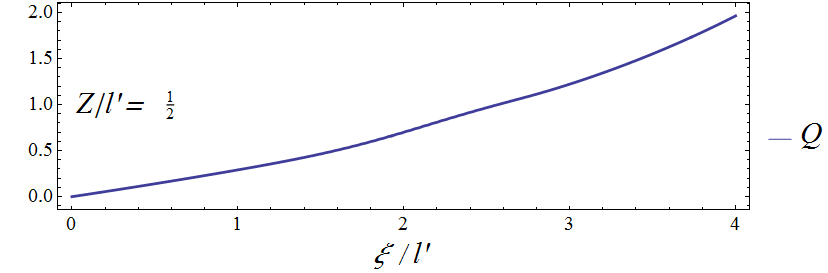}
\end{minipage}%
\hfill
\begin{minipage}{.49\textwidth}
\includegraphics[width=7.3cm]{InitialDensityPlot_Nonlin}\\
\includegraphics[width=8cm]{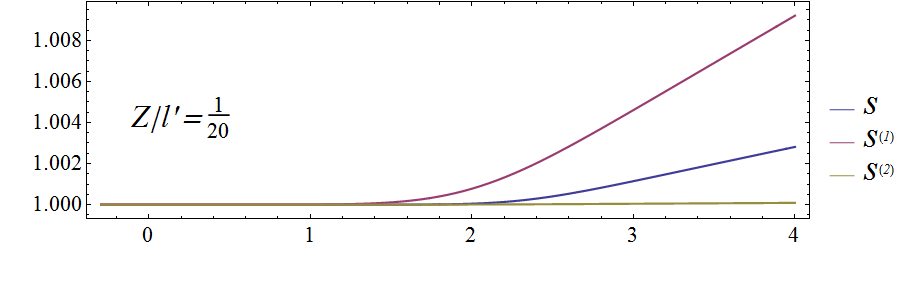}\\
\includegraphics[width=8cm]{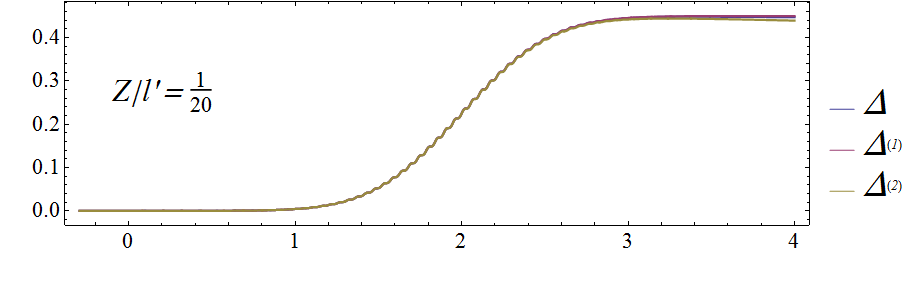}\\
\includegraphics[width=7.9cm]{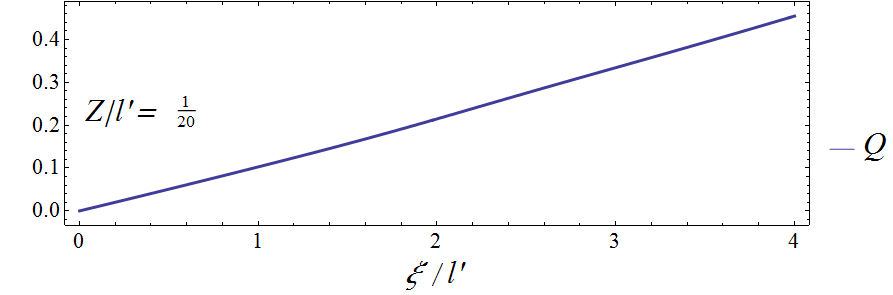}\\
\includegraphics[width=8cm]{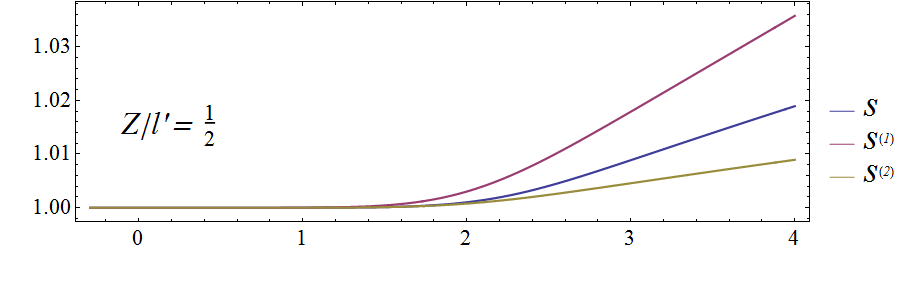}\\
\includegraphics[width=8cm]{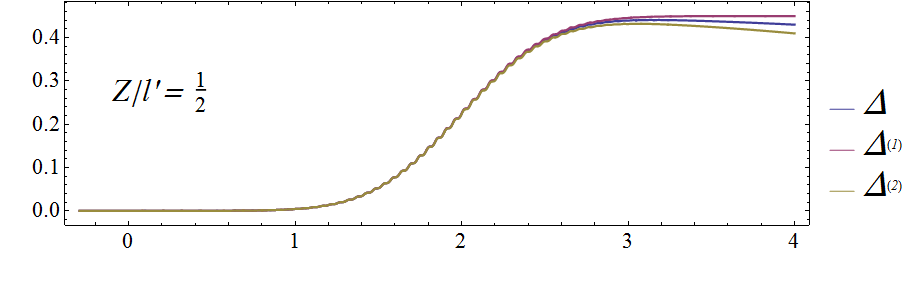}\\
\includegraphics[width=7.9cm]{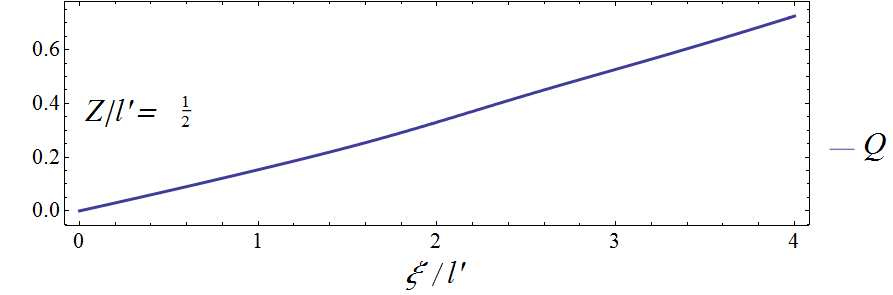}
\end{minipage}
\caption{%Up to down:  
The initial electron densities 3), 4) of fig. \ref{InDensityPlots} (respectively left, right) 
with $n_b=2\times 10^{18}$cm$^{-3}$, and 
corresponding plots of $\hs,\hD$,  their upper and lower bounds
$\hsU,\hsD,\hDU,\hDD$, and the function $Q(\xi,Z)$, vs. $\xi$ for the same sample
values  $Z=l'/20$   and  $Z=l'/2$ of $Z$. The values $Q_2(Z) \!:=\! Q(l,\!Z)$ can be read off the plots. 
As we can see, the bounds are much better for the density 4); 
the values $Q_2(Z) \le 1$ are consistent
with all worldlines intersecting rather far from the laser-plasma interaction spacetime region. 
Whereas the large value of $Q_2(Z)$ for  the density 3)
is an indication that some worldlines intersect  not far from  the laser-plasma interaction spacetime region.
Our computations lead also to $Q_0(l'/20)=4.62$, $Q_0(l'/2)=3.08$ with the density 3),
 $Q_0(l'/20)=0.49$,  $Q_0(l'/2)=0.85$ with the density 4).}
\label{graphs-bounds_Homlin-Nonlin}
\end{figure}

\begin{figure}[t]
\includegraphics[width=17.0cm]{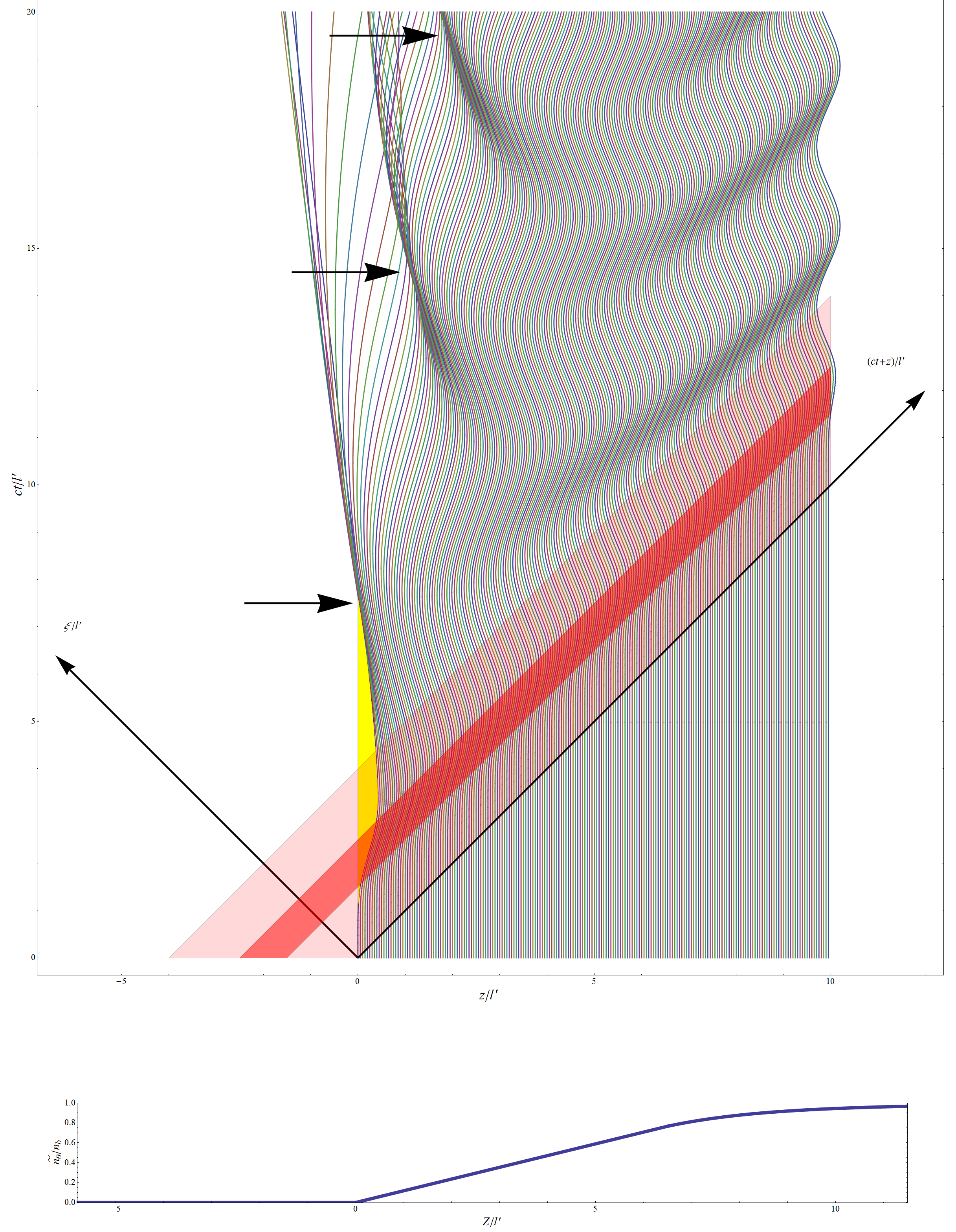}
\caption{Down:  the initial electron density 3)   of fig. \ref{InDensityPlots}.
Up:  The worldlines of $Z$-electrons interacting with the  pulse of fig. \ref{graphs}  for 200 equidistant values of $Z$;
  the support $0\le ct\!-\!z\le l$ and the 'effective support' $(l\!-\!l')/2\le ct\!-\!z\le (l\!+\!l')/2$ of the pulse are pink, red; the spacetime region of the pure-ion layer is yellow. Horizontal arrows pinpoint where particular subsets of worldlines first intersect.}
\label{Worldlines_HomLin-Nonlin}
\end{figure}

\begin{figure}[t]
\includegraphics[width=17.5cm]{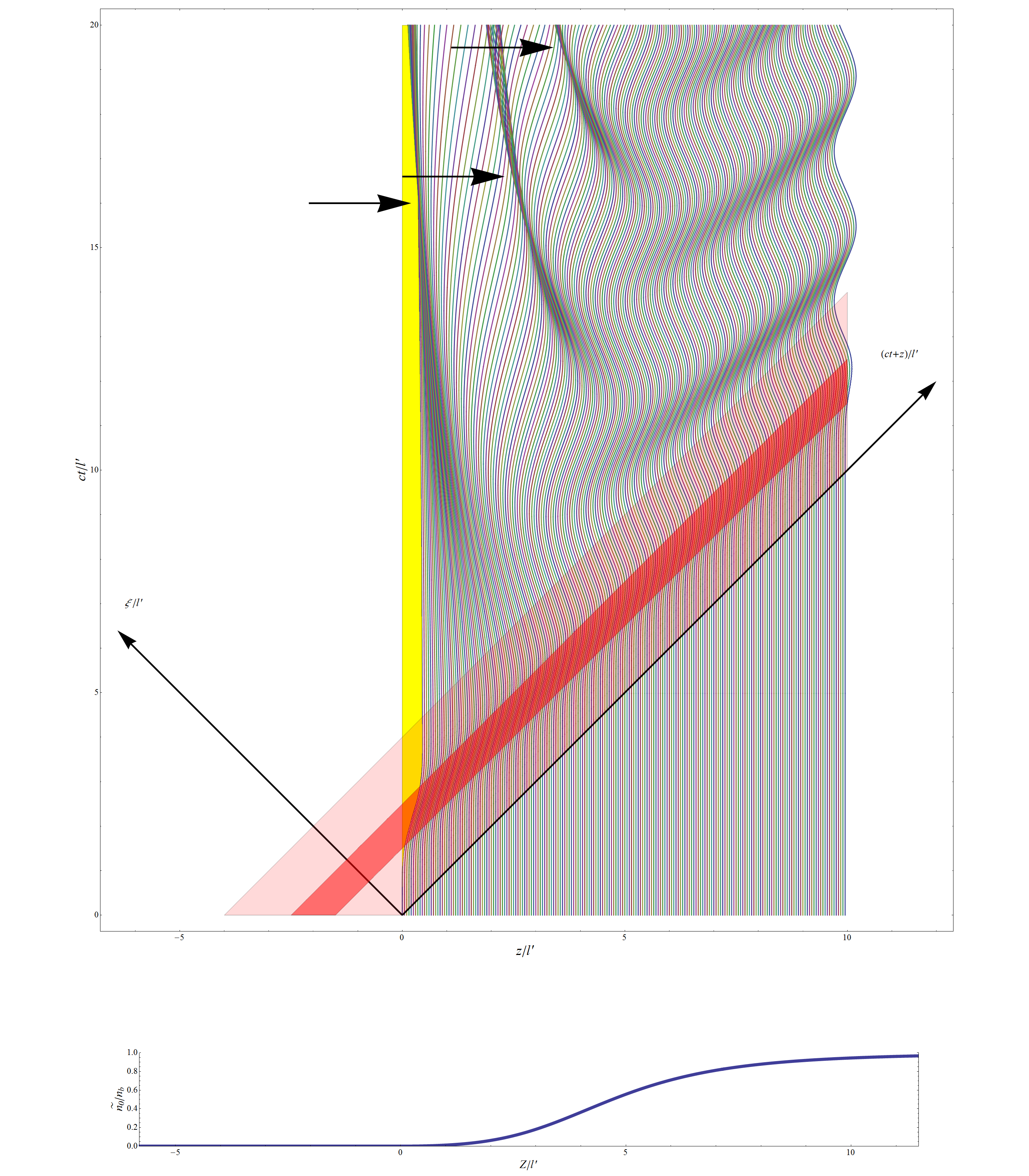}
\caption{Down:    the initial electron density 4)  of fig. \ref{InDensityPlots}.
Up:  The worldlines of $Z$-electrons interacting with the  pulse of fig. \ref{graphs}  for 200 equidistant values of $Z$;
  the support $0\le ct\!-\!z\le l$ and the 'effective support' $(l\!-\!l')/2\le ct\!-\!z\le (l\!+\!l')/2$ of the pulse are pink, red; the spacetime region of the pure-ion layer is yellow. Horizontal arrows pinpoint  where particular subsets of worldlines first intersect; as we can see,  small $Z$ worldlines first intersect quite farther from the laser-plasma interaction spacetime region (in pink)
than in the linear homogenous case 3).}
\label{Worldlines_Nonlin}
\end{figure}

The upper bound  $n_b$ plays also the role of asymptotic value.
Here we choose  $n_b=2\times 10^{18}$cm$^{-3}$, which is the same as the $n_0$ of fig.  \ref{graphs}, 
what yields  $Kn_bl'{}^2\simeq 4$;{}
but the results  for the dimensionless variables remain the same if we change $n_b,\lambda, l'$
keeping $\lambda/l'$ and $Kn_bl'{}^2$ constant.
As already said, in the case of the step-shaped density 0), if  $\tilde\xi_2> l$, i.e. if
the pulse is strictly short [a sufficient condition for that is  (\ref{ShortPulse1})], then $n_u=n_d=n_b$, $\delta=0$,  and by
(\ref{NoWBcond2}-\ref{NoWBcond0}) there is no  WBDLPI; more directly, this is a consequence of $J(\xi,Z)\equiv 1$ for $0\le\xi\le l$, which  follows from the $Z$-independence of $\hD$ in such an interval.
As for the other profiles, we have   respectively plotted:

\begin{itemize} 

\item In fig. \ref{J-sigma-dudz_Homlin-LinInhom-const-1}
and fig. \ref{J-sigma-dudz_Homlin-Nonlin} the solutions $J,\sigma$ of eq. (\ref{basic1}) and, by (\ref{Def-dudz}), the associated function $l'\partial u^z/\partial Z$  for $0\le \xi\le 6l'$ and a few
values of $Z$, assuming the initial electron density profiles 1), 2) and 3), 4)  respectively. 

\item  In fig. \ref{graphs-bounds_Homlin-LinInhom-const-1}
and fig. \ref{graphs-bounds_Homlin-Nonlin} the solutions $\hs,\hD$ of (\ref{heq1}-\ref{incond}), their upper and lower bounds
$\hsU,\hsD,\hDU,\hDD$  [eq. (\ref{sDelta1Def})] and the function $Q$ of eq. (\ref{NoWBcond2}),  for $0\le \xi\le 6l'$ and a few
values of $Z$, assuming the initial electron density profiles 1) - 2) and 3) - 4), respectively. 

\item In fig. \ref{Worldlines_HomLin-Nonlin} and fig. \ref{Worldlines_Nonlin} the corresponding worldlines
of the $Z$-electrons, for $0\le ct\le 20l'$ and $Z=n\, l'/20$, $n=0,1,...,200$,  associated to the initial electron density profiles  3), 4). The  support of the EM pulse is coloured pink (the red part is the more intense part); the laser-plasma interaction takes place in the  spacetime region which has nonempty intersection with worldlines of electrons or protons.

\end{itemize} 
We can compare   the results for the densities 1) - 2)
and those for the densities 3) - 4) side-by-side. 
In case 1) WBDLPI is avoided assuming that $\widetilde{n_0}(0)>0$,
but worldlines intersect and WB  takes place not  far from the  laser-plasma interaction region; in 
case 2) WBDLPI takes place for $Z\simeq 0$, due to the steep growth of
$\widetilde{n_0}(Z)=Z/l'n_b$ from the value $\widetilde{n_0}(0)=0$. 
In case 3), albeit  the growth $\widetilde{n_0}(Z)\propto Z$ is much less steep, again
worldlines intersect and WB  takes place not very far from the  laser-plasma interaction region, whereas in case
4) this occurs quite far from the latter, consistently with  the results $|J-1|\ll1$, $Q_2<1$.
Thus we note that, although such dynamics are moderately relativistic, rather than nonrelativistic, switching from profile 1) to profile 2), or from profile   3) to profile 4) has the same qualitative effect of avoiding (or distancing from) WBDLPI.
We also note that in case 1), 3)  the $Z\simeq 0$ worldlines first  intersect with very small angles, or equivalently, that when the corresponding electrons collide their longitudinal momenta differ by a very small amount: by   formula (\ref{Def-dudz}) 
$l \partial u^z/\partial Z$ is very small  not only because $\sigma$ is, but also because $v\simeq 0$ and $\hs>1$. Hence we can expect that these collisions will lead only to a very small momentum spreading.

Essentially the same results are reached choosing a different pulse polarization, because $v_a$ will be of the same type. In the case of circular polarization ($\varphi_1\!-\!\varphi_2=\pm \pi/2$, $\cos\psi=\pm\sin\psi$)
and Gaussian modulation $v$ itself will essentially coincide with $v_a$, thus displaying a single maximum (cf. fig. \ref{graphs}b).

%%pausa%%

\bigskip
As mentioned in the introduction, the equations of motion  for the $Z$-electrons
can be reduced to the form (\ref{heq1}), and thus
are {\it decoupled} from those of all $Z'$-electrons, $Z'\neq Z$, only 
in the idealization where the laser pulse is  "undepleted", i.e. not affected  by its interaction
with the plasma. The latter is expected to be an acceptable approximation 
 only for small $t,z>0$. Actually, for a slowly modulated monochromatic wave one can show by self-consistency \cite{Fio-impact} that
this is a good approximation in the spacetime region that is the intersection of the  `laser-plasma interaction' stripe
\ $0\le ct-z\le l$ \ with the orthogonal stripe
\be
0 \:  \le  \: %\frac {M_b}{8\pi}\lambda (ct+z)\equiv
\frac{e^2 n_b\lambda}{2 mc^2}(ct+z)
%=\frac{\pi e^2 n_b}{k mc^2}(ct+z)
 \: \ll  \: 1.
\label{neglectDepletion}
\ee
In view of the inequalities $\lambda\ll l$ and (\ref{Lncond}) or  (\ref{Lncond'})  we see that,
to our satisfaction, this region is much longer than $l$ in  the $ct\!+\!z$ direction.
 Thus, if $n_b=2\times 10^{18}$cm$^{-3}$, and the pulse is as in fig. \ref{graphs}, we can consider the latter as undepleted, and the electrons' motion determined above as accurate,
for time intervals $[0,t_d]$, where $t_d$ is at least a few $10^{-13}$s.

\bigskip
The above  predictions  are based on idealizing the initial laser pulse as a plane EM wave
(\ref{pump}). In a more realistic
picture the $t=0$ laser pulse is cylindrically symmetric around the $\vec{z}$-axis and has a {\it finite} spot radius $R$, namely the $t=0$ EM fields are of the form
$\bE=\Be^{{\scriptscriptstyle\perp}}\!(-z)\,\chi(\rho),\:\: \bB=
\bk\!\times\!\bE$,
 where $\rho^2=x^2\!+\!y^2$ and $\chi(\rho)\ge 0$ is 1 for $\rho\le 1$ and rapidly goes to zero for $\rho>R$. By causality, the motion of the electrons   remains \cite{FioDeN16} strictly the same in the  
future Cauchy development\footnote{We recall that the  {\it future Cauchy development} $D^+(\D)$  of a  region $\D$ in Minkowski spacetime $M^4$ is defined as the set of all points $x\in M^4$ for which every past-directed causal (i.e. non-spacelike)  
line through $x$  intersects $\D$.} $D^+(\D)$ 
of $\D=\D_1\cup\D_2$, where $\D_1:=\{(ct,\bx)=(0,0,0,z>0\}$ and $\D_2:=\{(0,\bx) \,|\, \rho\le R\}$,
 and almost the same in a neighbourhood of $D^+(\D)$;
therefore the conditions described above remain sufficient to exclude WBDLPI at least in such a region. 

\medskip
Finally, the conditions  of Proposition \ref{PropNoWB} are very general, in that they apply
also to discontinuous  $\widetilde{n_0}$, or non-monotone  $\widetilde{n_0}$, but  if  $\widetilde{n_0}$ has a bounded
derivative it turns out that they are unnecessarily  too strong for ensuring that no WBDLPI occurs.
Weaker no-WBDLPI conditions under the latter assumptions will be treated in \cite{Fio-impact}.

%\bigskip
%\noindent
%{\bf Funding:} \ This research received no external funding. 

%\bigskip
%\noindent
%{\bf Conflicts of interest:} \ The authors declare no conflict of interest.

\section{Appendix}

\subsection{Proof of Proposition \ref{propshort}}
\label{Proofpropshort}

We abbreviate $\Omega:=\sqrt{Kn_b}$, $\zeta:=l-\eta$. Eq. (\ref{LncondNR}) yields
\bea
2\Delta(l) &=& \int_0^{l/2}\!\!\!\!d\eta\,v(\eta)\,\cos\left[\Omega(l\!-\!\eta)\right]+
\int^l_{l/2}\!\!\!\!d\eta\,v(l\!-\!\eta)\,\cos\left[\Omega(l\!-\!\eta)\right]\nn
  &=& \int_0^{l/2}\!\!\!\!d\eta\,v(\eta)\,\cos\left[\Omega(l\!-\!\eta)\right]-
\int^0_{l/2}\!\!\!\!d\zeta\,v(\zeta)\,\cos\left(\Omega\zeta\right)\nn
  &=& \int_0^{l/2}\!\!\!\!d\eta\,v(\eta)\,\cos\left[\Omega(l\!-\!\eta)\right]+
\int^{l/2}_0\!\!\!\!d\eta\,v(\eta)\,\cos\left(\Omega\eta\right)\nn
  &=& \left[1\!+\!\cos\left(\Omega l\right)\right]\int_0^{l/2}\!\!\!\!d\eta\,v(\eta)\cos\left(\Omega\eta\right)+
\sin\left(\Omega l\right)\int_0^{l/2}\!\!\!\!d\eta\,v(\eta)\sin\left(\Omega\eta\right)  \label{Dl}\\
2\delta(l)  &=& \int_0^{l/2}\!\!\!\!d\eta\,v(\eta)\,\sin\left[\Omega(l\!-\!\eta)\right]+
\int^l_{l/2}\!\!\!\!d\eta\,v(l\!-\!\eta)\,\sin\left[\Omega(l\!-\!\eta)\right]\nn
&=& \int_0^{l/2}\!\!\!\!d\eta\,v(\eta)\,\sin\left[\Omega(l\!-\!\eta)\right]-
\int^0_{l/2}\!\!\!\!d\zeta\,v(\zeta)\,\sin\left(\Omega\zeta\right)\nn
  &=& \left[1\!-\!\cos\left(\Omega l\right)\right]\int_0^{l/2}\!\!\!\!d\eta\,v(\eta)\sin\left(\Omega\eta\right)+
\sin\left(\Omega l\right)\int_0^{l/2}\!\!\!\!d\eta\,v(\eta)\cos\left(\Omega\eta\right) \label{dl}
\eea
If $\Omega l\le\pi$ both integrands in (\ref{Dl}) are nonnegative, and so are the factors of
both integrals; moreover the latter and $\Delta(l)$ itself are positive if $\Omega l<\pi$, zero if $\Omega l=\pi$. In either case  $\Delta(\xi)>0$ for all $\xi\in]0,l[$, because $\Delta'(l)=v(l)-\delta(l)<0$ (since $v(l)\simeq 0$, $\delta(l)>0$). Moreover, if
$\varepsilon:=\Omega l\!-\!\pi>0$ is sufficiently small, then both integrals and $1\!+\!\cos\left(\Omega l\right)=1\!-\!\cos\varepsilon\simeq \varepsilon^2/2$ are still positive,  whereas $\sin\left(\Omega l\right)=-\sin \varepsilon\simeq-\varepsilon$; the second negative term will dominate and make (\ref{Dl})  negative as well.
Therefore (\ref{LncondNR}a) will be satisfied iff $\Omega l\le\pi$ i.e. if $G_b\le 1/2$.

Similarly, if $\Omega l=2\pi$ the factors of both integrals in (\ref{dl}) vanish, and $\delta l=0$. 
If $\varepsilon:=\Omega l\!-\!2\pi\neq 0$ is sufficiently small, then  $1\!-\!\cos\left(\Omega l\right)=1\!-\!\cos\varepsilon\simeq \varepsilon^2/2$,  whereas $\sin\left(\Omega l\right)=\sin \varepsilon\simeq \varepsilon$, and the second  term dominates over the first. Under our assumptions the second integral will be negative, because $v(\eta)$ is larger where $\cos\left(\Omega\eta\right) <0$. Hence (\ref{dl}) will be  negative if $\varepsilon>0$ (and  sufficiently small); if $\varepsilon<0$ (and sufficiently small), then  (\ref{dl}) will be  positive, and $\delta(\xi)>0$
also for $0<\xi<l$, because $\delta'(l)=M\Delta(l)<0$ (since in this case $\Delta(l)<0$.
Therefore (\ref{LncondNR}b) will be satisfied iff $\Omega l\le2\pi$ i.e. if $G_b\le 1$.

\subsection{Estimates of oscillatory integrals}
\label{oscill}

Here we recall some useful estimates \cite{Fio18JPA} %\CITE{FioCat19} 
of oscillatory integrals, such as (\ref{pump2}a) in the case  (\ref{modulate}).
Given a  function $f\in C^2(\mathbb{R})$, % (the Schwartz space), 
integrating by parts we find  for all $n\in\mathbb{N}$
\bea
&&\int^\xi_{-\infty}\!\!\!\!\!\! d\zeta\: f(\zeta)e^{ik\zeta} = -\frac ik f(\xi)e^{ik\xi}+R_1^f(\xi), \label{modula'}\\
&&  R_1^f(\xi) := \frac ik \int^\xi_{-\infty}\!\!\!\!\!\! d\zeta\: f'(\zeta)\,e^{ik\zeta}=\left(\frac ik\right)^2\left[-f'(\xi)\,e^{ik\xi}
+\!\int^\xi_{-\infty}\!\!\!\!\!\! d\zeta\:  f''(\zeta)\, e^{ik\zeta}\right] .  
     \label{Rnf}
\eea
Hence we find the following upper bounds for the remainder \ $ R_1^f$:
\bea
&& \displaystyle\left| R_1^f(\xi) \right|
\le \frac {1}{|k|^2}\left[ |f'(\xi)|+\displaystyle \int^\xi_{-\infty}\!\!\!\!\!\! d\zeta\, |f''(\zeta)|\right]
 \le \frac { \Vert f' \Vert_\infty+ \Vert f''\Vert_1}{|k|^2},
%\quad\Rightarrow\quad R_1^f\!=\!O\left(\frac 1{k^2}\right)
\qquad  \label{oscineqs1}
\eea
It follows $R_1^f\!=\!O(1/k^2)$. All inequalities in (\ref{oscineqs1}) are useful:
the left inequalities are more stringent, while the right ones are $\xi$-independent.

Equations (\ref{modula'}), (\ref{oscineqs1}) and $R_1^f\!=\!O(1/k^2)$ hold also if $f\in W^{2,1}(\mathbb{R})$ (a Sobolev space), in particular
if $f\in C^2(\mathbb{R})$ and $f,f',f''\in L^1(\mathbb{R})$, because the previous steps can be done also under such assumptions. Equations (\ref{modula'}) will hold with a remainder $R_1^f\!=\!O(1/k^2)$ also under weaker assumptions, 
e.g. if  $f'$ is bounded and piecewise continuous and $f,f',f''\in L^1(\mathbb{R})$, but $R_1^f$ will be
a sum of contributions like  (\ref{Rnf}) for every interval in which $f'$ is continuous.

\noindent
Letting $\xi\!\to\!\infty$ in (\ref{modula'}), (\ref{oscineqs1}) we find   for the Fourier  transform 
$\tilde f(k)=\displaystyle\int^{\infty}_{-\infty}\!\!\!\!\!\! d\zeta\,  f(\zeta)e^{-iky}$  of $f(\xi)$
\be
|\tilde f(k)|\le  \frac { \Vert f' \Vert_\infty+ \Vert f''\Vert_1}{|k|^2},
\ee 
hence  $\tilde f(k)=O(1/k^2)$ as well.
Actually, if $f\!\in\!{\cal S}(\mathbb{R})$ then   $\tilde f(k)$ decays much faster as $|k|\to \infty$, 
since  $\tilde f\!\in\!{\cal S}(\mathbb{R})$ as well.
For instance, if $f(\xi)=\exp[-\xi^2/2\sigma]$ then $\tilde f(k)=\sqrt{\pi\sigma}\exp[-k^2\sigma/2]$.

To prove approximation (\ref{slowmodappr}) now we just need to choose $f=\epsilon$ and note that
every component of $\Bap$ will be a combination of (\ref{modula'})
and  (\ref{modula'})$_{k\mapsto -k}$.

%\bigskip
%{\bf Funding}: \ This research did not receive any specific grant from funding agencies in the public, commercial, or not-for-profit sectors.

\begin{thebibliography}{}


\bibitem{Kruer19} W. Kruer, 
{\it The Physics Of Laser Plasma Interactions},
%3rd edition,
CRC Press, 2019, 200 pp.
%ISBN: 9781000754209, 1000754200

\bibitem{SprEsaTin90PRA}
P. Sprangle,  E. Esarey,  A. Ting, 
Nonlinear interaction of intense laser pulses in plasmas, 
Phys. Rev. {\bf A41} (1990), 4463.

\bibitem{SprEsaTin90PRL}
P. Sprangle, E. Esarey,  A. Ting, 
Nonlinear Theory of Intense Laser-Plasma Interactions, 
Phys. Rev. Lett. {\bf 64} (1990), 2011.

\bibitem{EsaSchLee09}
E. Esarey, C. B. Schroeder,  W. P. Leemans,
{\it Physics of laser-driven plasma-based electron accelerators},
Rev. Mod. Phys.  {\bf 81} (2009), 1229.

\bibitem{Mac13}
A. Macchi, {\it A Superintense Laser-Plasma Interaction Theory Primer},
Springer, 2013. %, 114 pp.
%ISBN: 9789400761254, 9400761250

\bibitem{KuzRyz17}
 V.V. Kuzenov, S.V. Ryzhkov,  {\it Numerical simulation of the effect
of laser radiation on matter in an external magnetic field},
J. Phys. Conf. Ser. {\bf 830} (2017), 012124. 

\bibitem{Tajima-Dawson1979} T. Tajima, J.M. Dawson, 
Laser Electron Accelerator,
Phys.Rev.Lett. \textbf{43}, 267 (1979).

\bibitem{Sprangle1988} P. Sprangle, E. Esarey, A. Ting,  G. Joyce, 
Laser wakefield acceleration and relativistic optical guiding,
Appl. Phys. Lett. \textbf{53}, 2146 (1988).

\bibitem{TajNakMou17}
T. Tajima, K. Nakajima, G. Mourou,
{\it Laser acceleration}, 
Riv. N. Cim. {\bf 40} (2017),34.
%DOI 10.1393/ncr/i2017-10132-x

\bibitem{HidEtAl19}
Hidding B., et al., {\it Fundamentals and Applications of Hybrid
LWFA-PWFA}, Appl. Sci. {\bf 9} (2019),
2626; https://doi.org/10.3390/app9132626.

\bibitem{Eupraxia19AIP}
 M. K. Weikum,  et al.,
{\it EuPRAXIA – a compact, cost-efficient particle and radiation source},
AIP Conf. Proceedings 2160, 040012 (2019).
% https://doi.org/10.1063/1.5127692

\bibitem{Eupraxia19JPCS}
M. K. Weikum,  et al., {\it 
Status of the Horizon 2020 EuPRAXIA conceptual design study},
 J. Phys.: Conf. Ser.  {\bf 1350} (2019), 012059. \ 
%doi:10.1088/1742-6596/1350/1/012059


\bibitem{Eupraxia20EPJ}
  R. W. Assmann, et al., {\it EuPRAXIA Conceptual Design Report}, 
Eur. Phys. J.: Spec. Top., {\bf 229} (2020), 
%Issue 24, December 2020, Pages 
3675-4284; {\it Erratum to: EuPRAXIA Conceptual Design Report}, 
Eur. Phys. J.: Spec. Top., {\bf 229} (2020), 
%Issue 24, December 2020, Pages 
4285-4287.

\bibitem{TomEtAl17}
P. Tomassini, S. De Nicola, L. Labate, P. Londrillo, R. Fedele, D. Terzani, L. A. Gizzi, 
{\it  The resonant multi-pulse ionization injection},  
Phys. Plasmas 24, 103120 (2017). % https://doi.org/10.1063/1.5000696

\bibitem{AkhPol56}
A. I. Akhiezer, R. V. Polovin, 
{\it Theory of wave motion of an electron plasma},
Sov. Phys. JETP {\bf  3}, 696 (1956).

\bibitem{GorKir1987} 
L.M. Gorbunov,  V.I. Kirsanov, 
{\it Excitation of plasma waves by an electromagnetic wave packet},
%Russian original - Zh. Eksp. Teor. Fiz. 93, 509 (1987) % (ZhETF, Vol. 93, No. 2, p. 509),
Sov. Phys. JETP \textbf{66} (1987), 290.

\bibitem{RosBreKat91}
J. Rosenzweig, B. Breizman,  T. Katsouleas, J. Su, 
{\it Acceleration and focusing of electrons in two-dimensional nonlinear plasma wake fields},
Phys. Rev.  {\bf A44} (1991), R6189.

\bibitem{MorAnt96}
P. Mora, T. M. Antonsen,  
\textit{Electron cavitation and acceleration in the wake of an ultraintense, self-focused laser pulse},
Phys. Rev. {\bf  E53} (1996), R2068(R).
%https://doi.org/10.1103/PhysRevE.53.R2068


\bibitem{PukMey2002}
A. Pukhov, J. Meyer-ter-Vehn,
{\it Laser wake field acceleration: the highly non-linear broken-wave regime},
Appl. Phys. {\bf B74} (2002), pp 355–361.

\bibitem{KosPukKis2004} 
I. Kostyukov,  A. Pukhov, S. Kiselev, 
{\it Phenomenological theory of laser-plasma interaction in ‘bubble’ regime}, 
Phys. Plasmas {\bf 11} (2004), 5256.

\bibitem{LuEtAl2006} 
W. Lu, C. Huang,  M.Zhou, W. Mori,  T. Katsouleas, 
{\it Nonlinear theory for relativistic plasma wakefields in the blowout regime}, 
Phys. Rev. Lett., {\bf 96} (2006), 165002.

\bibitem{LuHuaZhoEtAl06}
W. Lu,  C. Huang, M. Zhou, M. Tzoufras, F. S. Tsung, W. B. Mori, T. Katsouleas, 
{\it A nonlinear theory for multidimensional relativistic plasma wave wakefields}, 
Phys. Plasmas 13 (2006), 056709.

\bibitem{MaslovEtAl16}
V.I. Maslov, O.M. Svystun, I.N.Onishchenko,V.I.Tkachenko,
{\it Dynamics of electron bunches at the laser–plasma interaction in the bubble regime},
 Nucl. Instr. Meth. Phys. Res. {\bf A829} (2016), 422.


\bibitem{FioFedDeA14}  G. Fiore, R. Fedele, U. de Angelis, 
{\it The slingshot effect: a possible new laser-driven high energy acceleration mechanism for electrons}, 
Phys.  Plasmas {\bf 21} (2014), 113105.
% doi: 10.1063/1.4901285. arXiv:1309.1400. 

\bibitem{FioDeN16} 
G. Fiore,  S. De Nicola,
{\it A simple model of the slingshot effect}, 
Phys Rev. Acc. Beams {\bf 19} (2016), 071302 (15pp).
%arXiv:1509.04656.

\bibitem{FioCat18} G. Fiore, P. Catelan,  
\emph{On cold diluted plasmas hit by short laser pulses},
Nucl. Inst. Meth. Phys. Res., {\bf A 909} (2018), 41-45.
%arXiv:1803.02915
% Proceedings of the "3rd European Advanced Accelerator Concepts (EAAC) Workshop", 
%24-30 September 2017, La Biodola, Isola d'Elba.  
%https://doi.org/10.1016/j.nima.2018.03.038

\bibitem{Fio-impact}
G. Fiore, T. Akhter,   S. De Nicola, R. Fedele, D.  Jovanovi\'c, 
\emph{On the impact of short laser pulses on cold diluted plasmas},
in preparation.

\bibitem{Fio14JPA} G. Fiore,   et al.,
\emph{On plane-wave relativistic electrodynamics in plasmas and in vacuum},  
J. Phys. A: Math. Theor. \textbf{47} (2014), 225501.
%doi:10.1088/1751-8113/47/22/225501
%arXiv:1312.4665 preprint,

\bibitem{Fio18JPA} G. Fiore,
\emph{Travelling waves and a fruitful `time' reparametrization in relativistic electrodynamics},  
J. Phys. A: Math. Theor. {\bf 51} (2018), 085203. 
%\ DOI: 10.1088/1751-8121/aaa304
%arXiv:1607.03482.
% https://doi.org/10.1088/1751-8121/aaa304

\bibitem{Fio14} G. Fiore,
\emph{On plane waves in Diluted Relativistic Cold Plasmas},  
Acta Appl. Math. \textbf{132}   (2014),  261. 
%261-271. %doi:10.1007/s10440-014-9901-4
% arXiv:1405.0163

\bibitem{Fio16b} G. Fiore,  
\emph{On very short and intense laser-plasma interactions},  
Ric. Mat.  {\bf 65} (2016), 491.
%491-503.  
%DOI 10.1007/s11587-016-0270-3. arXiv:1607.02367

\bibitem{FioCat19}  G. Fiore, P. Catelan, \ 
\emph{Travelling waves and light-front approach in relativistic electrodynamics},
Ric. Mat.  {\bf 68}  (2019), 341-357. 
%https://doi.org/10.1007/s11587-018-0411-y
%ISSN: 0035-5038
%arXiv:1807.00667

\bibitem{Fio21JPCS} G. Fiore,  
{\it Light-front approach to relativistic electrodynamics}, \
J. Phys.: Conf. Ser. 
%Journal of Physics: Conference Series, 
{\bf 1730} (2021), 012106.
%DOI: 10.1088/1742-6596/1730/1/012106. 
%Proceedings of the 9th International Conference on Mathematical Modeling in Physical Sciences (IC-MSQUARE), 7-10 September 2020, Tinos island, Greece
%https://iopscience.iop.org/article/10.1088/1742-6596/1730/1/012106

\bibitem{Daw59} J.D. Dawson, 
\textit{Nonlinear electron oscillations in a cold plasma},
Phys.Rev.{\bf 113} (1959),383

\bibitem{JovFedBelDeN19}
D. Jovanovic, R. Fedele, M. Belic, and S. De Nicola, 
{\it Adiabatic Vlasov theory of ultrastrong femtosecond laser pulse propagation in plasma. The scaling of ultrarelativistic quasi-stationary states: spikes, peakons, and bubbles},
Phys. Plasmas, {\bf 26} (2019), 123104. 


\bibitem{Yos66} T. Yoshizawa, {\it Stability Theory by Liapunov's second
 method}, Math. Soc. of Japan, 1966.

\bibitem{VeiEtAl11}
% Laszlo Veisz, Alexander Buck, Maria Nicolai, Karl Schmid, Chris M. S. Sears, Alexander Sävert, Julia M. Mikhailova, Ferenc Krausz, and Malte C. Kaluza, 
L.  Veisz, {\it et al.},  {\it Complete characterization of laser wakefield acceleration}, 
Proceedings SPIE 8079
%, Laser Acceleration of Electrons, Protons, and Ions; and Medical Applications of Laser-Generated Secondary Sources of Radiation and Particles, 
 (Prague, 2011), 807906
%Proc. SPIE 8079, Laser Acceleration of Electrons, Protons, and Ions; and Medical Applications of Laser-Generated Secondary Sources of Radiation and Particles, 807906 (25 May 2011); https://doi.org/10.1117/12.890952.

\bibitem{HosEtAl02}
%    T. Hosokai, K. Kinoshita, T. Watanabe, K. Yoshii, T. Ueda, A. Zhidokov, M. Uesaka, K. Nakajima, M. Kando, H. Kotaki,
 T. Hosokai, {\it et al.}, 
{\it Supersonic gas jet target for generation of relativistic electrons with 12-TW 50-fs laser pulse},
Particle accelerator. Proceedings, %8th European Conference, 
EPAC 2002 (Paris), %France, June 3-7, 2002, 
981-983.

\end{thebibliography}
\end{document}